\documentclass[journal]{IEEEtran}
\usepackage{graphicx, nicefrac,subfigure,amsfonts, amsmath,amssymb,color,array,booktabs}
\usepackage{algorithm,algpseudocode,cite,epstopdf,enumitem,url}
\graphicspath{{}}

\allowdisplaybreaks

\def\tX{\pmb{\mathcal{X}}}

\def\Re{\text{Re}}

\def\bbe{\boldsymbol{\beta}}

\def\bxi{\boldsymbol{\xi}}

\def\bPi{\boldsymbol{\Pi}}

\def\bLa{\mathbf{\Lambda}}

\def\bPh{\mathbf{\Phi}}
\def\bSi{\mathbf{\Sigma}}

\def\bPh{\mathbf{\Phi}}

\def\({\left(}
\def\){\right)}
\def\[{\left[\,}
\def\]{\,\right]}

\def\0{\boldsymbol{0}}
\def\1{\boldsymbol{1}}
\def\a{\mathbf{a}}
\def\A{\mathbf{A}}

\def\B{\mathbf{B}}

\def\c{\mathbf{c}}
\def\C{\mathbf{C}}
\def\bC{\mathbb{C}}

\def\D{\mathbf{D}}

\def\e{\mathbf{e}}
\def\E{\mathbf{E}}

\def\g{\mathbf{g}}
\def\G{\mathbf{G}}
\def\h{\mathbf{h}}
\def\H{\mathbf{H}}

\def\I{\mathbf{I}}

\def\J{\mathbf{J}}

\def\n{\mathbf{n}}

\def\P{\mathbf{P}}

\def\Q{\mathbf{Q}}

\def\s{\mathbf{s}}
\def\S{\mathbf{S}}
\def\t{\mathbf{t}}

\def\U{\mathbf{U}}

\def\v{\mathbf{v}}

\def\x{\mathbf{x}}
\def\X{\mathbf{X}}

\def\Y{\mathbf{Y}}
\def\z{\mathbf{z}}
\def\Z{\mathbf{Z}}

\def\diag{\mathrm{diag}}

\newtheorem{theorem}{Theorem}
\newtheorem{lemma}{Lemma}

\newtheorem{corollary}{Corollary}
\newtheorem{definition}{Definition}

\def\cO{\mathcal{O}}

\title{Tensor-Based Channel Estimation for Dual-Polarized Massive MIMO Systems}
\author{Cheng Qian, \emph{Member, IEEE,} Xiao Fu, \emph{Member, IEEE,} Nicholas D. Sidiropoulos, \emph{Fellow, IEEE}, Ye Yang
	\thanks{
		Preliminary conference version of part of this work will be presented at ICASSP 2018 \cite{icassp}. This work was supported in part by the National Science Foundation under project NSF ECCS 1808159 and NSF ECCS 1608961. \par
		C. Qian and N. D. Sidiropoulos are with the Department of Electrical and Computer Engineering, University of Virginia, Charlottesville, VA 22904 USA (e-mail: alextoqc@gmail.com, nikos@virginia.edu).\par
		X. Fu is with the School of Electrical Engineering and Computer Science, Oregon State University, Corvallis, OR 97331 (xiao.fu@oregonstate.edu). \par
		Y. Yang is with Physical Layer \& RRM IC Algorithm Department, WN Huawei Co., Ltd, Shanghai, China (yangye@huawei.com).\par
		MATLAB code is available at \protect\url{https://www.mathworks.com/matlabcentral/fileexchange/69176-tensor-based-channel-estimation-for-dual-polarized-mimo}
	}	
}

\begin{document}
	
	\maketitle
	
\begin{abstract}
	The 3GPP suggests to combine dual polarized (DP) antenna arrays with the double directional (DD) channel model for downlink channel estimation. This combination strikes a good balance between high-capacity communications and parsimonious channel modeling, and also brings limited feedback schemes for downlink channel state information within reach---since such channel can be fully characterized by several key parameters. However, most existing channel estimation work under the DD model has not yet considered DP arrays, perhaps because of the complex array manifold and the resulting difficulty in algorithm design. In this paper, we first reveal that the DD channel with DP arrays at the transmitter and receiver can be naturally modeled as a low-rank tensor, and thus the key parameters of the channel can be effectively estimated via tensor decomposition algorithms. 
	On the theory side, we show that the DD-DP parameters are identifiable under mild conditions, by leveraging identifiability of low-rank tensors. Furthermore, a compressed tensor decomposition algorithm is developed for alleviating the downlink training overhead. We show that, by using judiciously designed pilot structure, the channel parameters are still guaranteed to be identified via the compressed tensor decomposition formulation even when the size of the pilot sequence is much smaller than what is needed for conventional channel identification methods, such as linear least squares and matched filtering.
	Extensive simulations are employed to showcase the effectiveness of the proposed method.
\end{abstract}


\section{Introduction}
The dual-polarized (DP) antenna array has many appealing features that make it a strong candidate for adoption in next generation communication systems and massive MIMO \cite{polarization2,3gpp,3gpp_stand,polarization5}.
For example, Foschini and Gans \cite{polarization4} showed that the capacity of systems with DP antennas at the transmitter can be increased up to 50\% compared to systems without polarization. Besides the increased capacity, DP antenna arrays have other key advantages relative to single-polarization counterparts with the same number of antennas, including smaller form factor and easier installation, better interference mitigation capability, and higher link reliability.

In the recent releases of technical specifications suggested by the 3GPP, the DP array and the double directional (DD) channel model are considered key techniques \cite{polarization5,3gpp,3gpp_stand}. 
The DD channel model is parsimonious for multipath channels with a small number of dominant paths, and such scenarios arise in millimeter wave (mmWave) based wireless communications. Modlel parsimony is really essential for designing limited feedback schemes for downlink channel state information in frequency-division duplex (FDD) massive MIMO \cite{3gpp}. Specifically, 3GPP suggests that the mobile users estimate the DD channel parameters such as directions-of-arrival (DOAs), directions-of-departure (DODs), the complex path-loss associated with each path, and then feed back these parameters to the base station (BS). This strategy is rather economical, as it is expected that the number of dominant paths will be small to moderate in practical deployments.
On the other hand, there are very few works related to the DD-DP channel/parameter estimation problem. Most of the existing channel estimation algorithms such as \cite{limitedfeedback2,jarvis,jomp,panos,fangjun,fangjun2} do not take polarization into consideration, and thus cannot be applied to this particular kind of system.

There are many challenges in the way of estimating the key parameters of the DD-DP channel. First, considering polarization adds another level of difficulty on top of the (DODs, DOAs, path losses) parametrization,  which is already not easy to handle in some cases, e.g., when we have small-size pilot matrices or a large number of multipaths. Formulating the parameter estimation problem for the DD-DP channel in a mathematically tractable form and tackling it using effective signal processing tools that provide analytical performance guarantees is quite nontrivial. Second, although the DD-DP channel (or to be more precise, blocks of the channel) can be modeled using long-existing array processing models (as we will show), it is hard to apply the classic array processing algorithms (e.g., MUSIC \cite{music} and ESPRIT \cite{esprit}) for estimating the key parameters. The reason is that classic array processing methods usually work under relatively restrictive assumptions---e.g., MUSIC and ESPRIT need the number of multipaths to be smaller than the number of transmit and the number of receive antennas, which may not be satisfied in practice. Real systems often have to deal with more multipaths than antennas on one end of the link. Third, the conventional estimation methods use matched filtering or linear least squares to extract an estimate of the channel matrix out of the received signals, and then perform parameter identification. To do this, the pilot sequence has to be quasi-orthogonal or at least full row-rank, respectively. This is very expensive for massive MIMO systems if the number of transmit antennas is large.

Very recently, Zhu \emph{et. al.,} proposed an interesting framework for two-dimensional DOA and DOD estimation of wideband massive MIMO-OFDM systems with DP arrays \cite{polarization5}. The key idea behind this algorithm is to exploit a so-called multi-layer reference signal structure to estimate the arrival and departure angles. Specifically, the transmitter and receiver communicate with each other iteratively, and in each iteration the transmitter (or receiver) fixes a beam and then the receiver (or transmitter) varies a paired beamforming vector to receive (or transmit) data. This way, a closed-form formula for DOA/DOD can be asymptotically derived. The total number of iterations is proportional to the product of the number of paired beams and the number of radio frequency bins at both transmitter and receiver. This closed-loop iterative protocol implicitly assumes that the transmitter, receiver, and scatterers remain static, and the two ends of the link are synchronized in beam sweeping. Its resolution is also limited by the utilized beamwidth.

In this work, we consider the parameter estimation problem for frequency division duplex (FDD) dual-polarized DD channels---but the algorithms can be easily implemented in the time division duplex (TDD) systems as well. 
We aim at designing novel efficient channel estimation algorithms and analyzing the identifiability of the key parameters, i.e., DOAs, DODs and path-losses, of this model.
Unlike the existing methods for DD-DP channels as in \cite{polarization5}, our proposed approach does not require multiple iterations between the transmitter and receiver, which may not be desired or realistic in practice, e.g., in a scenario with relatively higher mobility. 
Our method is also naturally with high spatial resolution, inheriting nice properties of related array processing techniques.
In addition, we fully characterize the theoretical boundaries of our methods in terms of parameter identifiability, leveraging advanced tensor algebra, and show that our method can work under a variety of challenging scenarios where existing methods tend to fail.
Our detailed contributions can be summarized as follows:

\begin{itemize}[leftmargin=3mm]
	\item {\bf Tensor-Based Formulation.} 
	We show that the DD-DP channel can be naturally modeled as a low-rank tensor. Leveraging this structure, we recast the associated channel estimation problem as a low-rank tensor decomposition problem \cite{Sid2017} and handle it using effective tensor decomposition algorithms. 

	\item {\bf Rigorous Identifiability Analysis.} 
    On the theory side, we show that the channel (i.e., the multipath parameters) are identifiable under very mild and practical conditions---even when the number of paths largely exceeds the number of receive antennas, a practically important case that classic DP array processing algorithms, e.g., \cite{polarized1}, cannot cope with. 

    \item {\bf Reduced-pilot Formulation and Identifiability.} 
    We propose a downlink signaling strategy that utilizes a judiciously designed pilot structure. We show that this pilot structure combined with {\it compressed tensor modeling} can substantially reduce the downlink overhead, without losing identifiability of the channel parameters. This design is particularly suitable for massive MIMO systems, for which existing methods usually need very long pilot sequences to help the receivers extract the channel matrix and then perform parameter estimation. An effective estimation algorithm is also proposed for the designed piloting strategy.

\end{itemize}

We should mention that some very recent work \cite{fangjun,fangjun2} also studied the downlink and uplink channel estimation problems from a tensor decomposition viewpoint. Nevertheless, the work in \cite{fangjun,fangjun2} did not consider dual-polarized antenna arrays. Hence, the formulated problems and analyses there are quite different from ours.


A preliminary conference version of part of this work was presented at ICASSP 2018 \cite{icassp}. The conference version includes the basic modeling and identifiability claims without detailed proofs. This journal version additionally includes detailed proofs of the identifiability results, and the compressed tensor factorization formulation, its identifiability proof, and a new algorithm that handles the compressed formulation.


\noindent {\bf Notation:} Throughout the paper, superscripts $(\cdot)^T$, $(\cdot)^*$, $(\cdot)^H$, $(\cdot)^{-1}$ and $(\cdot)^\dagger$ represent transpose, complex conjugate, Hermitian transpose, matrix inverse and pseudo inverse, respectively. We use $|\cdot|$, $\|\cdot\|_F$, $\|\cdot\|_1$ and $\|\cdot\|_2$ for absolute value, Frobenius norm, $\ell_1$-norm and $\ell_2$-norm, respectively; $\hat a$ denotes an estimate of $a$, $\text{diag}(\cdot)$ is a diagonal matrix holding the argument in its diagonal, $\text{vec}(\cdot)$ is the vectorization operator and $\angle(\cdot)$ takes the phase of its argument; $[\cdot]_i$ is the $i$th element of a vector, $[\X]_{i,j}$ is the $(i,j)$ entry of $\X$, and $\x_{r,k}$ is the $k$th column of $\X_r$. Symbols $\otimes,\odot, \circledast \text{ and } \circ$ denote the Kronecker, Khatri-Rao, element-wise, and outer products, respectively; $[\X]_{[i:j,m:n]}$ extracts the elements in rows $i$ to $j$ and columns $m$ to $n$, $[\X]_{:,i:j}$ extracts the elements in the columns $i$ to $j$ and $[\X]_{i:j,:}$ extracts the elements in the rows $i$ to $j$. $\I_m$ is the $m\times m$ identity matrix and $\0_{m\times n}$ is the $m\times n$ zero matrix.

\section{Signal Model and Problem Statement}
\begin{figure}
	\centering
	\includegraphics[width=1\linewidth]{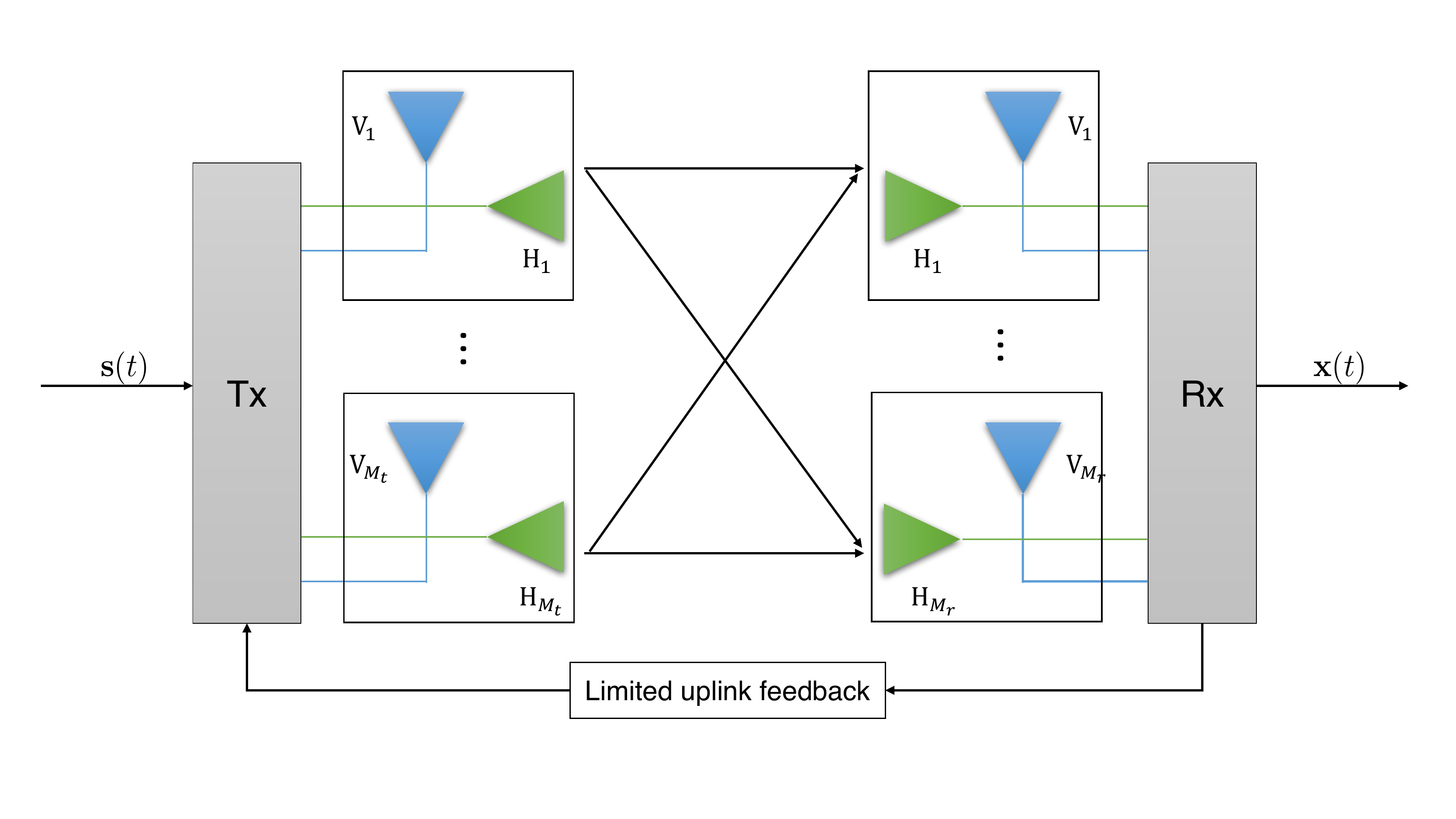}
	\vspace{-2.5em}
	\caption{DP MIMO system.}
	\label{fig:dpmimo}
\end{figure}

\subsection{Double Directional Dual-Polarized Channel Model}
We consider an FDD massive MIMO system, where there are $M_t$ DP transmit antennas and $M_r$ DP receive antennas, see Fig. \ref{fig:dpmimo}. In the  array processing literature, this type of DP array is also known as ``cross-polarized'' array \cite{polarized1}. 
Under the considered scenario, each antenna pair consists of a vertical (V) polarized antenna and a twin horizontal (H) polarized antenna---and each antenna is connected with an RF chain. 
Therefore, the channel can be represented as a $2M_r\times 2M_t$ matrix, where each element of the channel matrix represents a link between a transmit antenna (which could be a V-polarized or a H-polarized antenna) and a (V- or H-polarized) receive antenna.  
The signal received by the user is given by \cite{3gpp_stand}
\begin{align}\label{eq:channel}
\x(t) = \H\s(t) + \n(t),\ t=1,\cdots,N
\end{align}
where $\s(t)\in\bC^{2M_t \times 1}$ is the transmitted signal, $\n(t)$ is zero-mean i.i.d. circularly symmetric complex Gaussian noise.
The downlink channel matrix can be represented as 
\begin{align}\label{H}
\H = \begin{bmatrix} \H^\mathrm{(V_r,V_t)} & \H^\mathrm{(V_r,H_t)} \\ \H^\mathrm{(H_r,V_t)} & \H^\mathrm{(H_r,H_t)}  \end{bmatrix} \in\bC^{2M_r\times 2M_t},
\end{align}
where $\H^\mathrm{(V_r,V_t)}\in\bC^{M_r\times M_t}$ is a channel matrix between all the V-polarized transmit antennas and V-polarized receive antennas, $\H^\mathrm{(V_r,H_t)}\in\bC^{M_r\times M_t}$ is a channel matrix between all the H-polarized transmit antennas and V-polarized receive antennas, and likewise for the other two blocks in \eqref{H}.


For notational simplicity, let ${\rm p}\in\{\mathrm{V_r,H_r}\}$ and ${\rm q}\in\{\mathrm{V_t,H_t}\}$. Then, according to the channel model suggested by the 3GPP \cite{3gpp_stand}, the $(\mathrm{p,q})$th subchannel matrix is modeled as
\begin{align}\label{Hpq}
\H^\mathrm{(p,q)} = \sqrt{\frac{\kappa}{\kappa+1}}\H_\mathrm{LOS}^\mathrm{(p,q)} + \sqrt{\frac{1}{\kappa+1}}\H_\mathrm{NLOS}^\mathrm{(p,q)}
\end{align}
where $\H_\mathrm{LOS}^\mathrm{(p,q)}$ is the component of line-of-sight (LOS) and $\H_\mathrm{NLOS}^\mathrm{(p,q)}$ is the component of non-line-of-sight (NLOS), $\sqrt{\frac{1}{\kappa+1}}$ and $\sqrt{\frac{\kappa}{\kappa+1}}$ are energy normalization factors with $\kappa$ being the ratio between the power related to the LOS and the power related to the NLOS, and
\begin{align}
\H_\mathrm{LOS}^\mathrm{(p,q)} &= \tilde{\beta}_1^\mathrm{(p,q)}\a_{r}(\theta_{1},\phi_{1})\a_{t}^H(\vartheta_{1},\varphi_{1}) \label{Hlos} \\
\H_\mathrm{NLOS}^\mathrm{(p,q)} &= \sum_{k=2}^{K} \tilde{\beta}_k^\mathrm{(p,q)}\a_{r}(\theta_{k},\phi_{k}) \a_{t}^H(\vartheta_{k},\varphi_{k}) \label{Hnlos}
\end{align}
where the first path is assumed to be the LOS path that usually exists in systems operating at millimeter wave (mmWave) frequencies, and $K$ is the number of paths between the two subarrays.
$\a_{r}(\theta_{k},\phi_{k})\in\mathbb{C}^{M_r}$ is associated with the array manifold of subarray ${\rm p}$: $\theta_k$ and $\phi_k$ are azimuth and elevation DOAs of the $k$th path, respectively. 
Similarly,  $\a_{t}(\vartheta_{k},\varphi_{k})\in\mathbb{C}^{M_t}$ is determined by the subarray ${\rm q}$, and $\vartheta_{k}$ and $\varphi_{k}$ are the azimuth and elevation DODs of the $k$th path, respectively.
Note that $\{\tilde{\beta}_k^\mathrm{(p,q)}\}$ are generalized path-losses, which are random variables and affected by the small-scale loss, large-scale loss, distance between BS and MS, and dual-polarization parameters. 
We may express
\begin{align}
\tilde{\beta}_k^\mathrm{(p,q)}=\alpha_k\tilde{\gamma}_k^\mathrm{(p,q)}, \quad k=1,\cdots, K
\end{align}
where $\alpha_k$ denotes the standard path-loss which is caused by propagation and fading, while $\tilde{\gamma}_k$ denotes the polarization factor. Without loss of generality, we can absorb $\sqrt{\frac{\kappa}{\kappa+1}}$ and $\sqrt{\frac{1}{\kappa+1}}$ into $\tilde{\gamma}_1^\mathrm{(p,q)}$ and $\{\tilde{\gamma}_k^\mathrm{(p,q)}\}_{k=2}^K$, respectively, and define $\gamma_1^\mathrm{(p,q)}=\sqrt{\frac{\kappa}{\kappa+1}}\tilde{\gamma}_1^\mathrm{(p,q)}$ and $\Big\{\gamma_k^\mathrm{(p,q)}=\sqrt{\frac{1}{\kappa+1}}\tilde{\gamma}_k^\mathrm{(p,q)}\Big\}_{k=2}^K$. Thus,
\begin{align}
\beta_k^\mathrm{(p,q)} = \alpha_k\gamma_k^\mathrm{(p,q)},\; k=1,\cdots,K.
\end{align}
Substituting \eqref{Hlos} and $\eqref{Hnlos}$ into \eqref{Hpq} produces
\begin{align}
\H^\mathrm{(p,q)} = \A_r\diag\big(\bbe^\mathrm{(p,q)}\big)\A_t^H
\end{align}
where
\begin{align}
\A_r &= \begin{bmatrix} \a_{r}(\theta_{1},\phi_{1}) & \cdots & \a_{r}(\theta_{K},\phi_{K}) \end{bmatrix} \\
\A_t &= \begin{bmatrix} \a_{t}(\vartheta_{1},\varphi_{1}) & \cdots & \a_{t}(\vartheta_{K},\varphi_{K}) \end{bmatrix} \\
\bbe^\mathrm{(p,q)} &=
\begin{bmatrix}
\beta_1^\mathrm{(p,q)} &  \cdots & \beta_K^\mathrm{(p,q)}
\end{bmatrix}^T.
\end{align}
Now the channel matrix in \eqref{H} can be rewritten as
\begin{align}\label{eq:3-Dmimo}
\H &= 
\begin{bmatrix}
\A_r & \\ & \A_r
\end{bmatrix}
\begin{bmatrix}
\diag\big(\bbe^\mathrm{(V_r,V_t)}\big) & \diag\big(\bbe^\mathrm{(V_r,H_t)}\big) \\
\diag\big(\bbe^\mathrm{(H_r,V_t)}\big) & \diag\big(\bbe^\mathrm{(H_r,H_t)}\big) 
\end{bmatrix} \times\notag\\
&\quad  \begin{bmatrix}
\A_t & \\ & \A_t
\end{bmatrix}^H
\end{align}
which in a more compact form is
\begin{align}\label{H3}
\H = (\I_2\otimes\A_r) \bLa (\I_2\otimes\A_t)^H
\end{align}
where
\begin{align}
\bLa = 
\begin{bmatrix}
\diag\big(\bbe^\mathrm{(V_r,V_t)}\big) & \diag\big(\bbe^\mathrm{(V_r,H_t)}\big) \\
\diag\big(\bbe^\mathrm{(H_r,V_t)}\big) & \diag\big(\bbe^\mathrm{(H_r,H_t)}\big) 
\end{bmatrix}.
\end{align}

The model in \eqref{eq:3-Dmimo} has been advocated by the 3GPP as a standardized channel modeling approach for the long-term evolution (LTE) systems \cite{3gpp_stand,3gpp}.
As mentioned in \cite{3gpp_stand}, most standardized channels like spatial channel model (SCM), SCM extension (SCME) \cite{scme}, WINNER \cite{winner} and ITU \cite{itu} are based on this model.
The model assumes that the H-polarized subarray and the V-polarized subarray share the same array manifolds, while the polarization information is contained in the path-loss vectors, i.e., ${\bbe}^{( {\rm p},{\rm q})}$'s. This model has many favorable features. It concisely models the effect of polarization. More importantly, it incorporates the elevation information of the transmit and receive antenna arrays in addition to the azimuth information---leading to the so-called 3-D channel modeling, which is considered very useful for next generation wireless communication systems, since it provides many more degrees of freedom that can potentially enhance system performance; see detailed discussion in \cite{3gpp_stand,3gpp}.

\begin{figure}
	\centering
	\includegraphics[width=0.7\linewidth]{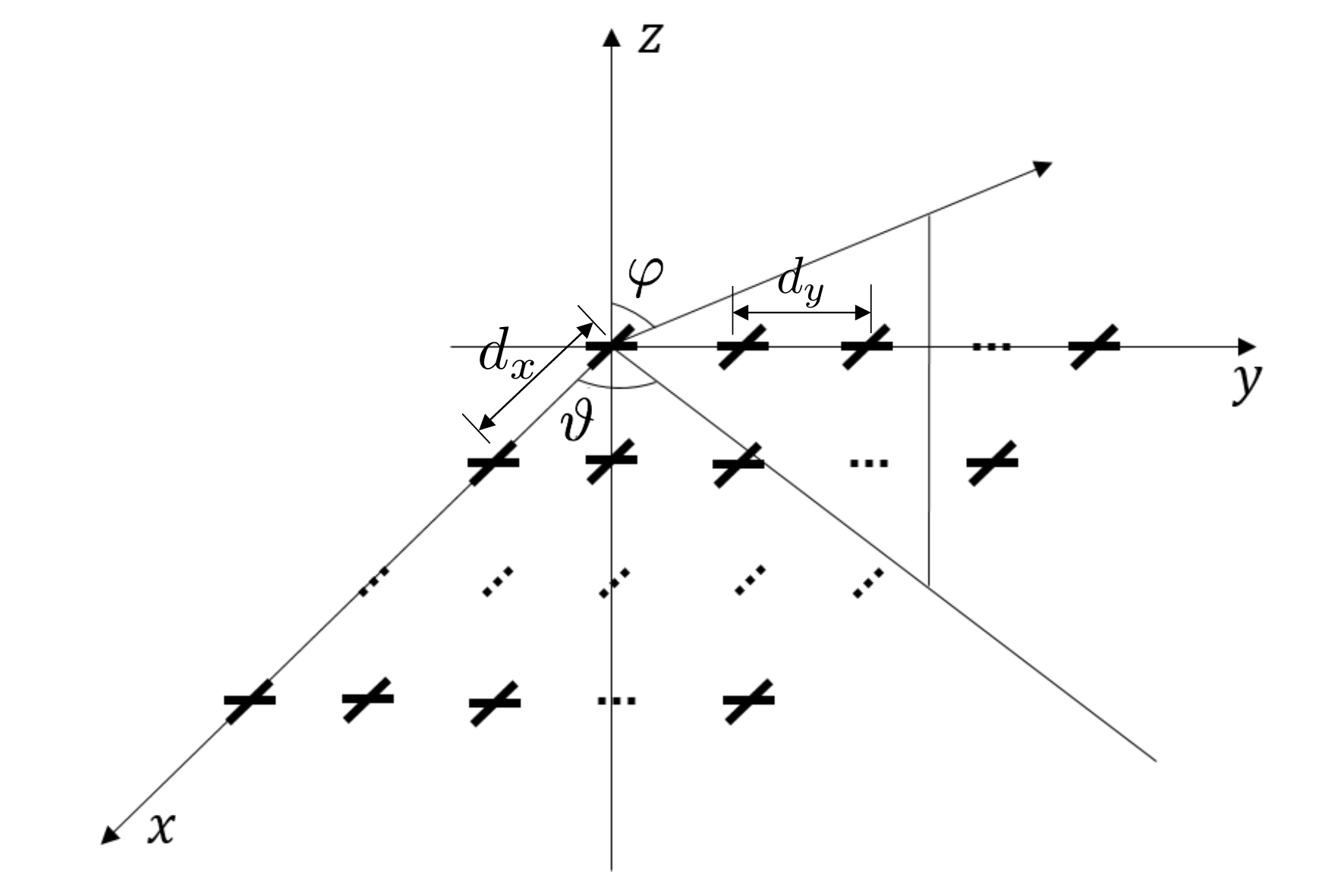}
	\caption{Illustration of a URA at the BS. Each ``cross'' represents a DP antenna pair.}
	\label{fig:ura}
\end{figure}

In practice, the specific form of $\a_{r}(\theta_{k},\phi_{k})$ and $\a_{t}(\vartheta_{k},\varphi_{k})$ are intimately tangled with the array geometry.
For example, the BSs are usually equipped with uniform rectangular arrays (URAs) that each has $M_x$ and $M_y$ horizontal and vertical array units\footnote{Note that in the DP arrays, each array unit consists of a pair of DP antennas.}. An illustration is shown in Fig.~\ref{fig:ura}. In this special case, the $k$th steering vector for the transmitter becomes
\begin{align}\label{eq:ura_manifold}
\a_t(\vartheta_{k},\varphi_{k})=\a_{t,k} = \a_{y,k} \otimes \a_{x,k}
\end{align}
where $[\a_{x,k}]_{l_x} = e^{j\omega_{x,k}}, l_x = 0,\cdots,M_x-1$ and $[\a_{y,k}]_{l_y} = e^{j\omega_{y,k}},l_y=0,\cdots,M_y-1$
with $\omega_{x,k}=2\pi (l_x-1)d_x\sin(\varphi_k)\cos(\vartheta_k)/\nu$ and $\omega_{y,k} = 2\pi (l_y-1)d_y\sin(\varphi_k)\sin(\vartheta_k)/\nu$. Here, $\nu$ is the wavelength, $d_x$ and $d_y$ are the inter-element spacing distances for horizontal and vertical units, respectively. 

For ease of exposition, let us assume that the receivers employ uniform linear arrays (ULAs).
Note that the analytical tools that we use can be easily extended to cover cases where the receive array has a different geometry, e.g., URA.
Under the ULA assumption, the $k$th steering vector for the receiver is
\begin{align}
\a_{r}(\theta_{k},\phi_{k})=\a_{r,k} = 
\begin{bmatrix}
1 & e^{j\omega_{r,k}} & \cdots & e^{j(M_r-1)\omega_{r,k}}
\end{bmatrix}^T
\end{align}
where $\omega_{r,k}=2\pi d_r\sin(\theta_k)/\nu$ with $d_r$ being the inter-element spacing of the ULA at the receiver side.
Note that to avoid phase wrapping, $d_x$ and $d_y$ must be carefully chosen. The most widely adopted choice for $d_x$ and $d_y$ is half-wavelength. 
In this paper, given the angular ranges of $\varphi$, $\vartheta$ and $\theta$, we assume that $d_x$, $d_y$ and $d_r$ are determined such that $2\pi d_x\sin(\varphi)\cos(\vartheta)/\nu \leq \pi$, $2\pi d_x\sin(\varphi)\sin(\vartheta)/\nu \leq \pi$ and $2\pi\sin(\theta)/\nu\leq \pi$ for all $ \varphi,\vartheta,\theta$ in their own ranges, respectively.

\subsection{Problem Statement}
Given the described channel model, our goal is to estimate the key parameters of the 3-D downlink channel.
Here, by ``key parameters'', we mean the set of DOAs and DODs that are associated with the paths and the corresponding path-losses, i.e., $\{\theta_k,\phi_k\}_{k=1}^K$, $\{\vartheta_{k},\varphi_{k}\}_{k=1}^K$ and ${\bbe}^{({\rm p},{\rm q})}$ for all ${\rm p}\in\{\mathrm{V_r,H_r}\}$ and ${\rm q}\in\{\mathrm{V_t,H_t}\}$.
Note that for massive MIMO systems that follow the channel model in \eqref{eq:3-Dmimo}, one is very well motivated to estimate these parameters. The reason is that the (potentially large) channel matrix ${\H}$ that has $4M_rM_t$ complex-valued elements is fully characterized by the parameters of interest. Since the number of multipaths is usually not large in practice, the number of parameters is relatively small, i.e., $8K$ ($2K$ for the DOAs, $2K$ for the DODs and $4K$ for the complex-valued path-losses) and we usually have
$$ 8K\ll 4M_rM_t. $$
For example, in a scenario where $M_r=M_t=20$ and $K=6$ paths exist, the channel matrix consists of $1,600$ complex-valued elements
while we only have 48 key parameters, of which $24$ are real-valued (DOAs and DODs).
Therefore, by estimating the key parameters, one can feedback the downlink channel to the BS in a very economical way---only feeding back the parameters rather than the whole channel ${\H}$ suffices to recover the downlink MIMO channel at the BS.
In fact, this is the main idea enabling implementation of limited feedback schemes in mmWave massive MIMO systems suggested by the 3GPP \cite{3gpp}.

In this work, we begin with the scenario where the downlink channel matrix ${\H}$ can be estimated at the receiver via simple procedures such as matched filtering or linear least squares (LS).
Our objective is to estimate the key parameters from a given ${\H}$.
We will first study identifiability theory associated with this simpler scenario---i.e., under what conditions the DOAs, DODs and path-losses can be provably identified from ${\H}$? Effective algorithms based on the identifiability analysis will also be proposed.
In addition, we will consider more challenging yet desirable scenarios where only a compressed version of ${\H}$ available at the receiver, and provide a pragmatic and effective algorithm to estimate the parameters of interest---this approach can substantially reduce the pilot length, thereby greatly saving the downlink training overhead.

\subsection{Prior Art and Challenges}
Estimating the DOAs, DODs and path-losses from ${\H}$ is not a trivial task.
Nevertheless, many classic methods from the array processing society can be applied, under some conditions.
For example, if the submatrix
$\H^\mathrm{(p,q)} = \A_r\diag\big(\bbe^\mathrm{(p,q)}\big)\A_t^H$ has full-column rank $\A_r$, $\diag\big(\bbe^\mathrm{(p,q)}\big)$ and $\A_t$, and also if the receive and transmit antenna arrays are ULAs/URAs, the subspace methods such as ESPRIT and MUSIC can be applied to estimate the DOAs and DODs.
After that, the path-losses can be recovered, e.g., via LS estimation.

This is viable, but can only work under relatively stringent conditions.
The classic array processing methods like ESPRIT and MUSIC all assume 
full column rank of $\A_r$, $\diag\big(\bbe^\mathrm{(p,q)}\big)$, and $\A_t$, which implicitly assumes that $K\leq \min\{M_r,M_t\}$ (to be precise, $K+1\leq \min\{M_r,M_t\}$ is needed for subspace methods like MUSIC).
In many practical scenarios, $M_r$ is relatively small---e.g., the newest model of iPhone (i.e., iPhone X released in 2017) only supports two receive antennas, while the number of paths can easily exceed two. Is there a more powerful method that can provably identify the parameters of interest under much more relaxed conditions? We will address this question in the next section.

Another possible way to handle the parameter estimation problem is to treat each block $\H^\mathrm{(p,q)} = \A_r\diag\big(\bbe^\mathrm{(p,q)}\big)\A_t^H$ as a sparse optimization problem \cite{jarvis,jomp}.
For each subchannel block, we discretize the DOA and DOD domains into fine angle grids and then construct  three overcomplete angle dictionaries (codebooks), denoted by $\D_r$, $\D_x$ and $\D_y$. Then, we have $\H^\mathrm{(p,q)}\approx \D_r\G^\mathrm{(p,q)}(\D_y\otimes\D_x)^H$, where $\G^\mathrm{(p,q)}$ is a sparse matrix that selects out the columns associated with the active DODs and DOAs from the dictionaries. This way, the parameter estimation problem becomes a sparse recovery problem that can be handled by formulations such as LASSO \cite{tibshirani1996regression}, i.e., $$\min_{\g^\mathrm{(p,q)}}~\|\h^\mathrm{(p,q)}-(\D_y^\ast\otimes\D_x^\ast \otimes {\bf D}_r) \g^\mathrm{(p,q)}\|_2^2 + \lambda \|\g^\mathrm{(p,q)}\|_1$$
where $\h^\mathrm{(p,q)}={\rm vec}(\H^\mathrm{(p,q)})$ with $\g^\mathrm{(p,q)}={\rm vec}(\G^\mathrm{(p,q)})$; and other sparse optimization algorithms such as orthogonal matching pursuit \cite{jomp}. 
The difficulty is that to ensure good spatial resolution, $\D_r\in\mathbb{C}^{M_r\times D_r}$, $\D_x\in\mathbb{C}^{M_x\times D_x}$ and $\D_y\in\mathbb{C}^{M_y\times D_y}$ are very ``fat'' matrices, where $D_r$, $D_x$ and $D_y$ denotes the number of angle grid points after quantization. Consequently, $( \D_y^\ast\otimes\D_x^\ast \otimes \D_r)$ is of size $M_rM_xM_y\times D_rD_xD_y$. If one quantizes the DOA and DOD space (ranging from $-90^\circ$ to $90^\circ$) using a resolution of one degree, then $D_rD_xD_y=5,929,741$---which poses a very hard sparse optimization problem.


\section{Proposed Approach}
In this section, we propose to estimate the key parameters of interest using low-rank tensor factorization---which has provable guarantees under realistic and relaxed conditions.

\subsection{Tensor Preliminaries}
To make the paper self-contained, we briefly present the definition of tensor and some useful theorems on the uniqueness of tensor decomposition in the following.

\begin{definition}\label{definition:tensor}
	\textit{(Tensor)}. A tensor is a multidimensional array indexed by three or more indices. Specifically, an $N$-th order tensor $\tX\in\bC^{I_1\times \cdots \times I_N}$ that has $N$ latent factor matrices $\U_1\ \cdots\U_N$ can be written as
	\begin{align*}
	\tX=\sum_{f=1}^{F} [\U_1]_{:,f} \circ \cdots \circ [\U_N]_{:,f}
	\end{align*}
	where $\U_n\in\bC^{I_n\times F}$ and the minimal such $F$ is the rank of tensor $\tX$ or the canonical polyadic decomposition (CPD) rank of $\tX$. \cite{Sid2017}.
\end{definition}

\begin{definition}\label{definition:unfolding}
(\textit{Unfolding}).	For an $N$-th order tensor $\tX\in\bC^{I_n\times \cdots \times I_N}$ in Definition \ref{definition:tensor}, its $n$-mode matrix unfolding can be written  as
	\begin{align*}
	\X_{(n)} =  (\U_N\odot\cdots\odot\U_{n+1}\odot\U_{n+1}\odot\cdots\odot\U_1) \U_n^T.
	\end{align*}
\end{definition}
Simply speaking, each unfolding is obtained by taking the mode-$n$ slabs of the tensor (i.e., subtensors obtained by fixing the $n$th index of the original tensor), vectorizing the slabs, and then stacking all the vectors into a matrix---see details in \cite{Sid2017}.

Low-rank tensor decomposition [also known as CPD or Parallel Factor Analysis (PARAFAC)] aims at factoring $\tX$
into a sum of column-wise outer products of $\U_1,\ldots,\U_N$---with each such outer product being a rank-one tensor. 
Unlike matrix factorization, which is in general non-unique, the PARAFAC decomposition has unique solution under mild conditions, up to scaling and permutation of the $F$ components. One of the best-known uniqueness results for third-order tensors is due to Kruskal \cite{kruskal}, which was later extended to higher orders by Sidiropoulos and Bro \cite{nikos1}.
\begin{theorem}\label{kruskal:uniqueness}
	\cite{nikos1} Given a $N$th order tensor as in Definition \ref{definition:tensor}, if $\sum_{n=1}^{N}{\rm krank}(\U_n)\geq 2F+N-1$, then $\mathrm{rank}(\tX)=F$ and the decomposition of $\tX$ is essentially unique, where ${\rm krank}(\cdot)$ denotes {\em Kruskal rank}.
\end{theorem}
The essential uniqueness makes the latent factors of a tensor identifiable from the `ambient data' $\tX$ up to some trivial ambiguities---which has enabled a tremendous amount of applications---see an overview in \cite{Sid2017}.
Theorem~\ref{kruskal:uniqueness} is known to be a broadly applicable general result.
In some special cases where the factor matrices have special structure, e.g., Vandermonde structure, the uniqueness condition in Theorem \ref{kruskal:uniqueness} can be improved. For example, we have the following theorem:
\begin{theorem}\cite{sid2001ident}
	Consider a tensor $\tX=\sum_{f=1}^{F} [\U_1]_{:,f} \circ [\U_2]_{:,f} \circ [\U_3]_{:,f}$, where $\U_1\in\mathbb{C}^{I_1\times F}$, ${\U_2}\in\mathbb{C}^{I_2\times F}$ and ${\U_3}\in\mathbb{C}^{I_3\times F}$ with $\U$ Vandermonde
	having distinct nonzero generators. Then, if
	$${\rm krank}({\U_2})+\min\{I_1 + {\rm krank}({\U_3}),2F\}\geq 2F+2 $$
	the factors are essentially unique.\label{theorem:sid2011ident}
\end{theorem}
Theorem~\ref{theorem:sid2011ident} presents a much milder identifiability condition relative to that in Theorem~\ref{kruskal:uniqueness}. The result is tailored for tensors that have a latent factor with Vandermonde structure. Such structure emerges quite often in array processing, since some array geometries (ULA, URA, nested or coprime arrays) 
naturally give rise to Vandermonde matrices.


\subsection{Tensor-Based Method and Identifiability}\label{section3a}
Our proposed approach starts by noticing that ${\H}$ is in fact a tensor of rank (at most) $K$, when the BS is equipped with a URA and the receiver with a ULA. 
To see this, let us first vectorize each block $\H^\mathrm{(p,q)}$ as
$\h^\mathrm{(p,q)} = \mathrm{vec}(\H^\mathrm{(p,q)}) = \left(\A_y^*\odot\A_x^*\odot\A_r\right)\bbe^\mathrm{(p,q)}$, and stack the vectorized $\H^\mathrm{(p,q)}$'s into a matrix as follows:
\begin{align}\label{H2}
\check{\H} &= \left[ \h^\mathrm{(V_r,V_t)}, \h^\mathrm{(V_r,H_t)}, \h^\mathrm{(H_r,V_t)}, \h^\mathrm{(H_r,H_t)} \right] \notag\\
& = \left(\A_y^*\odot\A_x^*\odot\A_r\right)\B^T
\end{align}
where
$\A_x=[\a_{x,1}\ \cdots\ \a_{x,K}]$ and $\A_y=[\a_{y,1}\ \cdots\ \a_{y,K}]$ and
\begin{align}
\B = [\bbe^\mathrm{(V_r,V_t)}\; \bbe^\mathrm{(V_r,H_t)}\; \bbe^\mathrm{(H_r,V_t)}\; \bbe^\mathrm{(H_r,H_t)}]^T \in\bC^{4\times K}
\end{align}
in which we have used \eqref{eq:ura_manifold} (or, more precisely $\A_t=\A_{x}\odot\A_y$) and ${\rm vec}({\X}{\rm diag}({\z)}{\Y}^H)=({\Y}^\ast\odot{\X}){\z}$.

Eq. \eqref{H2} is exactly the definition of a four-slab fourth-order tensor of rank $\leq K$ in the matrix unfolding form \cite{Sid2017} when $\min(M_r,M_x,M_y)>1$ (cf. Definition~\ref{definition:unfolding}).
We have this fourth-order tensor because of the array manifolds that we have assumed for the transmit and receive arrays.
The four factor matrices $\A_r$, ($\A_x$, $\A_y$) and $\B$ are the manifold of the receive antenna array, the manifold matrices of (horizontal and vertical) transmit antenna arrays, and the path-loss matrix, receptively.
A side comment is that if some other array geometries are employed at both sides, one can also derive a low-rank tensor from blocks of $\H$ by rearranging.
We list the resulting tensor structure of some pertinent cases for widely used configurations of the transmit and receive antenna arrays in Table I.

\begin{table}[t]\label{table1}
	\begin{center}
		\caption{Tensor order of $\H$ }
		\begin{tabular}{ ll|c } 
			\toprule
			\multicolumn{2}{c}{Array Configuration} & \multicolumn{1}{c}{Tensor order of $\H$}\\
			\midrule
			Tx & Rx & $M_r>1$  \\
			\midrule
			{DP-URA} & {DP-ULA} & {Four} \\ 
			{DP-URA} & {DP-URA} & {Five} \\
			DP-URA & DP-UCA & Four \\
			DP-UCA & DP-URA & Four \\
			\bottomrule
		\end{tabular}
	\end{center}
\end{table}

\subsubsection{Array Manifold Estimation}
Our idea is to estimate $\A_x$, $\A_y$, $\A_r$ and $\B$ from the tensor $\check{\H}$, and then estimate the multipath parameters using the estimated factor matrices.
As will be shown shortly, if $\A_x$, $\A_y$ and $\A_r$ are accurately estimated, the DOAs and DODs can then be estimated in closed-form.
To estimate the array manifolds and the path-losses (i.e., ${\B}$), we propose to employ the following tensor decomposition formulation:
\begin{align}\label{Prob:H1}
\min_{\A_r,\A_x,\A_y,\B} \left\| \check{\H} -  \left(\A_y^*\odot\A_x^*\odot\A_r\right)\B^T \right\|_F^2.
\end{align}
which is the least squares fitting formulation for low-rank tensor factorization.

Various low-rank tensor decomposition algorithms can be applied to identify the loading matrices \cite{parafac2,parafac3,Sid2017}. Among them, one of the most popular methods is the so-called alternating least squares (ALS) technique. To implement ALS for solving \eqref{Prob:H1}, we make use of different unfoldings of the tensor $\check{\H}$, which are denoted as follows:
\begin{align}
\H_{(1)} &= (\B\odot\A_y^*\odot\A_x^*) \A_r^T \\ 
\H_{(2)} &= (\B\odot\A_y^*\odot\A_r) \A_x^H \\
\H_{(3)} &= (\B\odot\A_x^*\odot\A_r) \A_y^H \\
\H_{(4)} &= (\A_y^*\odot\A_x^*\odot\A_r) \B^T.
\end{align}
Note that $\H_{(4)}$ is exactly $\check{\H}$.
Using the unfoldings, one can easily implement the following alternating optimization algorithm:
\begin{subequations}\label{eq:ten}
	\begin{align}
	\A_r &\leftarrow \arg\min_{\A_r}
	\left\|\H_{(1)}-(\B\odot\A_y^*\odot\A_x^* )\A_r^T\right\|_F^2 \label{sp1} \\
	\A_x &\leftarrow \arg\min_{\A_x}
	\left\|\H_{(2)}-(\B\odot\A_y^*\odot\A_r) \A_x^H\right\|_F^2 \label{sp2}\\
	\A_y &\leftarrow \arg\min_{\A_y}
	\left\|\H_{(3)}-(\B\odot\A_x^*\odot\A_r) \A_y^H\right\|_F^2 \label{sp3}\\
	\B &\leftarrow \arg\min_{\B}
	\left\|\H_{(4)}-(\A_y^*\odot\A_x^*\odot\A_r) \B^T\right\|_F^2 \label{sp4}
	\end{align}
\end{subequations}
where the four subproblems are all linear LS problems that can be readily solved in closed-form.
The ALS algorithm repeatedly solves the subproblems until convergence. Derivative-based schemes can also be used for optimization, from Gauss-Newton and BFGS to simple stochastic gradient type methods. See \cite{Sid2017} for more information.

\subsubsection{Parameter Estimation}

Once $\A_r$, $\A_x$, and $\A_r$ are obtained from \eqref{eq:ten}, the angles $\theta_k,\vartheta_{k}$ and $\varphi_{k}$ can be estimated by exploiting the manifold structure of $\A_{r,k},\A_{x,k}$ and $\A_{y,k}$.
To proceed, let us consider the following:
\begin{align}
\hat{\omega}_{r,k} &= \angle(\overline{\A}_{r,k}^H\underline{\A}_{r,k}) \label{omega_r}\\
\hat{\omega}_{x,k} &= \angle(\overline{\A}_{x,k}^H\underline{\A}_{x,k}) \label{omega_x}\\
\hat{\omega}_{y,k} &= \angle(\overline{\A}_{y,k}^H\underline{\A}_{y,k}) \label{omega_y}
\end{align}
where $\overline{\x}$ and $\underline{\x}$ are the vectors consisting of the first and last $(M-1)$ entries of $\x$ with length $M$, respectively. Then we estimate $\theta_k$ from $\hat{\omega}_{r,k}$, and $\vartheta_{k}$ and $\varphi_{k}$ from $\hat{\omega}_{x,k}$ and $\hat{\omega}_{y,k}$ as
\begin{align}
\hat{\theta}_{k} &= \sin^{-1}\left( \frac{\nu}{2\pi d_r}\hat\omega_{r,k} \right) \label{theta}\\
\hat{\varphi}_k &= \sin^{-1}\left( \sqrt{\left( \frac{\nu}{2\pi d_x}\hat{\omega}_{x,k}\right)^2 + \left( \frac{\nu}{2\pi d_y}\hat{\omega}_{y,k}\right)^2}  \right) \label{varphi}\\
\hat{\vartheta}_k &= \tan^{-1}\left( \frac{d_x\hat{\omega}_{y,k} }{d_y\hat{\omega}_{x,k}} \right). \label{vartheta}
\end{align}
Eqs.~\eqref{theta}-\eqref{vartheta} hold because we have assumed that the transmit and receive antenna arrays are URA and ULA, respectively. The closed-form solutions are also {\it rotationaly invariant}---not affected by scaling ambiguity that is brought by tensor decomposition. 

We should mention that the tensor decomposition algorithm in \eqref{eq:ten} has already given an initial estimate of ${\B}$, i.e., the path-losses.
However, since there is an intrinsic scaling ambiguity of tensor decomposition (cf. Lemma~1 in Appendix A), such an initial estimate may not be useful. Nevertheless, this is easy to fix. Note that $\hat{\A}_r,\hat{\A}_{x},\hat{\A}_y$ {\it  can be reconstructed} from $\theta_k,\vartheta_{k}$ and $\varphi_{k}$ without scaling ambiguity. 
Then, the estimate of ${\B}$ without scaling ambiguity can be computed as
\begin{align}\label{eq:recal_B}
	\hat{\B} &\leftarrow \arg\min_{\B}
\left\|\H_{(4)}-(\hat\A_y^*\odot\hat\A_x^*\odot\hat\A_r) \B^T\right\|_F^2.
\end{align}

Algorithm 1 summarizes the tensor based parameter estimation, where in the first step we have assumed that the pilot matrix ${\S}$ has orthogonal rows so that $\X\S^H$ gives a fairly accurate estimate of $\H$. Note that the order of applying tensor decomposition, angle estimation, and path-loss estimation matters---since angle decomposition can naturally remove the scaling ambiguity, as we discussed.

Note that the complexity of \eqref{omega_r}-\eqref{omega_y} is very low but the resulting estimate could be suboptimal. For better accuracy, one can resort to single-tone frequency estimation algorithms, e.g., \cite{freqest1,freqest2} or maximum likelihood (ML)-based (periodogram) methods, to estimate the DODs and DOAs from the estimated manifolds. These latter methods are statistically efficient (approximately) in the high SNR regime, but are computationally more demanding than the simple closed-form solutions provided earlier.

\subsubsection{Parameter Identifiability}
As we mentioned, the channel parameters can be estimated using some other methods, e.g., MUSIC and ESPRIT, which possibly admit lower complexity compared to tensor decomposition.
However, a salient feature of tensors is that the factors are uniquely identifiable under mild conditions. For example, it can be shown that $\{\A_r,\A_x,\A_y,\B\}$ meet the $k$-rank condition \cite{sid2001ident} provided that all the DOAs, DODs and path-losses are not the same, which is a mild condition considering the random nature of the multiple paths. Then we have the following theorem:
\begin{theorem}\label{theoremrank}
	Assume that the scenario where the transmitter is equipped with a URA and the receiver a ULA, 
	and that $(\theta_k,\vartheta_{k},\varphi_{k})$ and $(\theta_j,\vartheta_j,\varphi_j)$ are different for any $k\neq j$.
	Also assume that the pathloss parameters in $\B$ are generated following some jointly continuous distribution.
	Then, the key parameters $\{\theta_k,\vartheta_{k},\varphi_{k}\}_{k=1}^K$ and the path-losses ${\bbe}^{{\rm (p,q)}}$ for all ${\rm p}\in\{\mathrm{V_r,H_r}\}$ and ${\rm q}\in\{\mathrm{V_t,H_t}\}$ are uniquely identifiable via the proposed approach provided that
	\begin{align}\label{identifiability1}
	\min{(M_r,K)} + \min(M_x,K) &+ \min(M_y,K) \notag\\
	& + \min{(4,K)} \geq 2K + 3
	\end{align}
	almost surely.
\end{theorem}

The proof relies on the identifiability of the four-slab four-way tensors and is relegated to Appendix A.
Although the proof is relatively straightforward for someone who is frequently exposed to tensors,
the implication of Theorem~\ref{theoremrank} is important: Using the proposed approach, the key parameters are uniquely identifiable even when the number of paths largely exceeds the number of $\min\{M_r, M_t\}$. This makes the proposed method widely applicable to many realistic scenarios, especially where the receive antenna array is of a relatively small size, as in many mobile phones. Furthermore, this identifiability is independent of the array configurations of transmit and receive antennas.



Theorem~\ref{theoremrank} is intuitive and easy to read, but is not the best bound that we can get.
In fact, if we look at the parameter estimation problem from a multi-snapshot 2-D harmonic retrieval viewpoint, a much stronger identifiability result can be obtained, which is summarized in the following theorem:

\begin{theorem}\label{theorem2}
	The parameters $\big\{\theta_k,\varphi_k,\vartheta_k,\beta_k^{(\rm p,q)}\big\}$ are all uniquely identifiable provided that 
	\begin{align}
	K\leq \arg\max_{F,P_r,P_x,P_y} &~F\notag\\
	\mathrm{s. t.}\quad & \max\Big((P_r-1)P_xP_y, P_r(P_x-1)P_y, \notag\\
	&\quad\qquad P_rP_x(P_y-1)\Big) \geq F \notag\\
	& 8Q_rQ_xQ_y \geq F
	\end{align}
	where $P_r+Q_r=M_r+1, P_x+Q_x=M_x+1,P_y+Q_y=M_y+1$.
\end{theorem}

Theorem \ref{theorem2} can be proven by invoking identifiability results in multi-dimensional harmonic retrieval, in particular, the construction following the IMDF algorithm \cite{liu2}. The result in Theorem~\ref{theorem2} is a bit harder to read compared to Theorem~\ref{theoremrank}, but is far better upon close inspection. For example, when $M_x=4,M_y=8,\text{and }M_r=2$, the identifiable case under Theorem~\ref{theoremrank} is with $K$ up to $K=7$, while $K=32$ multi-paths can be guaranteed to be identified under Theorem \ref{theorem2}.
Furthermore, even when the MS only has a single dual-polarized antenna, it can be shown using the IMDF based approach that the number of identifiable paths is upper bounded by $K < 0.8187 M_t$. For details about the IMDF method, we refer the readers to \cite{liu2}.
Due to its simplicity, it is either a good candidate to initialize the method proposed in Section III-B or can be directly applied for computational efficiency.

\subsubsection{Computational Complexity}
We should remark that the complexity of the proposed tensor decomposition method is dominated by the step for solving \eqref{eq:ten}, where each subproblem is a least squares problem. Taking \eqref{sp1} as an example, the solution to this subproblem is as follows:
\begin{align*}
\hat{\A}_r^T =& \((\B^H\B)\circledast(\A_y^T\A_y^*)\circledast(\A_x^T\A_x^*)\)^{-1}\times \notag\\
&\(\B\odot\A_y^*\odot\A_x^*\)^H\H_{(1)},
\end{align*}
which needs $\cO\big( (4+M_x+M_y)K^2 + 4K^2+ K^3+4K^2M_xM_y + 4KM_rM_xM_y \big)$ flops if one uses the above relatively naive implementation.
The matrix inversion and large matrix product (i.e., $\(\B\odot\A_y^*\odot\A_x^*\)^H\H_{(1)}$) parts are the most costly to compute.
Nevertheless, these two operations can be avoided if some advanced solvers are employed, e.g., \cite{parafac2,xu2013block}.

\section{Parameter Identification Using Frugal Pilots}
In the previous section, we have proposed a tensor decomposition-based method for estimating the DOAs, DODs, and path-losses of the MIMO 3-D channel when a reliable estimate of ${\H}$ is available. This is viable when ${\S}$ is `fat' or square and with full row-rank. In other words, when the pilot sequence is long enough so that ${\S}$ has full row rank, ${\H}$ can be estimated via least squares (or simply matched filtering if $\S\S^H=\I$). Then, the method that is proposed in the previous section can be applied.
In practice, using a long pilot sequence is not desirable since this creates large downlink training overhead. When $M_t$ is large, the size of ${\S}$ is at least $2M_t\times 2M_t$ if one wishes to make the rows of ${\S}$ orthogonal to each other, that could be costly.
In this section, we propose another approach to handle the above challenge. We carefully design the transmit pilot sequence and formulate a \textit{compressed tensor decomposition (CTD)} problem for parameter estimation. As it turns out, we can use a pilot matrix whose size is much smaller than $2M_t \times 2M_t$ to identify the channel parameters.


\subsection{Proposed Downlink Training and Parameter Identification Approach}
To reduce downlink overhead in a massive MIMO system while keeping identifiability of the key parameters of interest, we propose to employ the following specially structured pilot matrix:
\begin{equation}\label{eq:pilot}
{\S}=\begin{bmatrix}
\Q&\0\\
\0&\Q
\end{bmatrix} \in \bC^{2M_t\times N}
\end{equation}
where $\Q\in\mathbb{R}^{M_t\times N/2}$ (assuming $N$ is an even number for simplicity) whose elements are generated following a certain absolutely continuous distribution and $N\in[4,M_t)$. 
This way, the (noise-free) received data matrix becomes

\begin{align}\label{eq:new_channel}
\X = \begin{bmatrix}
\A_r\diag\big(\bbe^\mathrm{(V_r,V_t)}\big)\A_t^H\Q & \A_r\diag\big(\bbe^\mathrm{(V_r,H_t)}\big)\A_t^H\Q\\
\A_r\diag\big(\bbe^\mathrm{(H_r,V_t)}\big)\A_t^H\Q & \A_r\diag\big(\bbe^\mathrm{(H_r,H_t)}\big)\A_t^H\Q
\end{bmatrix}
\end{align}
Given the above $\X$, our goal is to identify $\A_r$, $\A_t$, $\bbe^{\rm ( p,q)}$---i.e., the path losses and the array manifolds, since all the key parameters can be easily estimated from them as described in the previous section.

Physically, the proposed design of $\S$ in \eqref{eq:pilot} corresponds to a time-division multiplexing strategy that transmits pilots from the H-polarized array first, and then transmits the same pilots from the V-polarized array (or the other way around), with the other turned off---which is very easy to implement in practice. Nevertheless, as we will show, such a simple signaling strategy combined with tensor algebra allows us to identify the parameters of interest under very mild conditions---even when the number of columns of $\S$ is much smaller than $2M_t$.


\subsection{Identification Approach and Theoretical Guarantees}
One can see that the four blocks in \eqref{eq:new_channel} comprise a four-slab three-way tensor, where the $(\rm p,q)$th block is defined as
\begin{align}\label{Xi}
\X^{\rm (p,q)}=\A_r\diag\big(\bbe^\mathrm{(\rm p,q)}\big)\A_t^H\Q
\end{align}
for $\mathrm{p}\in\{\mathrm{V_r,H_r}\}$ and $\mathrm{q}\in\{\mathrm{V_t,H_t}\}$. Taking the transpose of $\X^{\rm (p,q)}$ and then vectorizing it yields
\begin{align}
\x^{\rm (p,q)}=\(\A_r\odot(\Q^T\A_t^*)\)\bbe^\mathrm{(\rm p,q)}.
\end{align}
Thus, by collecting $\{\x^{\rm (p,q)}\}$, we have
\begin{align}\label{Z}
\Z &= \[ \x^{\rm (V_r,V_t)} ~ \x^{\rm (V_r,H_t)} ~ \x^{\rm (H_r,V_t)} ~ \x^{\rm (H_r,H_t)} \] \notag\\
&= \(\A_r\odot\Q^T\A_t^*\)\B^T \in \bC^{NM_r/2\times 4}
\end{align}

It is readily seen that $\Z$ is nothing but a matrix unfolding of a third order tensor whose latent factors are $\Q^T\A_t^*$, $\B$ and $\A_r$. It immediately follows that $\Q^T\A_t^*$, $\B$ and $\A_r$ can be identified under the proposed pilot design under certain conditions.
Hence, at least the angle of arrivals can be easily estimated from $\hat{\A}_r$, under quite mild conditions that guarantee identifiability of third-order tensors, as stated in Theorem~\ref{kruskal:uniqueness}. Let
\begin{align}\label{E}
\E=(\Q^T\A_t^*)^*=\Q^H\A_t.
\end{align}
Estimating $\A_t$ from the compressed measurements $\E$ is nontrivial---can we uniquely identify the DODs from $\E$? The answer is affirmative---if the transmit array is an ULA or URA and certain mild conditions are satisfied, which we will explain shortly.

\subsubsection{Manifold Estimation via Smoothed ESPRIT}
There are many ways to estimate $\Q^T\A_t^*$, $\B$ and $\A_r$ from $\Z$, since this is nothing but a third-order tensor decomposition problem.
The identifiability of this kind of tensor (in which there is at least one factor which has a Vandermonde structure) has also been well understood---e.g., the aforementioned Theorem~2 that was derived in \cite{sid2001ident}. 
Here, we propose to employ a method that was recently proposed in \cite{smoothESPRIT} to handle this problem.
The method is in nature a subspace method, which works under very relaxed identifiability conditions by exploiting the Vandermonde structure of a latent factor of the tensor. 
The detailed proof of the algorithm can be found in \cite{smoothESPRIT}. Here, we refer to this algorithm as the \textit{smoothed ESPRIT} algorithm.

The algorithm starts by working with a third-order tensor as follows (with a bit of notational abuse):
\begin{align}\label{X}
\X = (\A\odot\B)\C^T
\end{align}
where  $\B$ and $\C$ are drawn from some continuous distributions, and $J\leq K$, and $\A\in\bC^{I\times F}$ is Vandermonde with distinct nonzero generators.
To identify $\A$, $\B$ and $\C$, one can employ the following procedure:

First, let us define a cyclic selection matrix $\J_{i_2}=[\0_{I_1\times i_2}~ \I_{I_1}~ \0_{I_1\times (I-i_2-I_1)}]$. It is easy to check that 
\begin{align}
(\J_{i_2}\otimes\I_J)\X &= \big((\J_{i_2}\A)\odot\B\big)\C^T \notag\\
&= (\A_1\otimes \B)\diag\big(\[\A\]_{i_2+1,:}\big)\C^T
\end{align}
where $\A_1=\J_{0}\A=[\A]_{1:I_1,:}$ contains the first $I_1$ rows of $\A$.
Thus, by varying $i_2$ from 0 to $(I_2-1)$ where $I_2=I+1-I_1$, we can construct a smoothed matrix as
\begin{align}\label{smoothedX}
\X_s &= 
\begin{bmatrix}
\J_{0} \X & \cdots & \J_{I_2-1} \X 
\end{bmatrix} \notag\\
&= \(\A_{1}\odot\B\) \( \A_2\odot \C \)^T \in \bC^{I_1J\times I_2K}.
\end{align}
where $\A_2$ takes the first $I_2$ rows of $\A$.
Starting from \eqref{smoothedX}, the smoothed ESPRIT algorithm applies a series of re-arranging of the tensor elements and finally converts the factorization problem to a classical DOA estimation problem that can be handled by ESPRIT---and solves it via eigen-decomposition---see details in \cite{smoothESPRIT}.
The algorithm also offers very favorable identifiablity guarantees:

\begin{theorem} \cite{smoothESPRIT}\label{theorem:3wayVdmd}
	Consider a third-order tensor $\X = \A (\B \odot \C)^T$, where $\A\in\bC^{I\times F}$, $\B\in\bC^{J\times F}$, $\C\in\bC^{K\times F}$, $\A$ is Vandermonde with distinct nonzero generators. Assume that $\B$ and $\C$ are drawn from certain absolutely continuous distributions, respectively. Then, if
	\begin{align}
	F \leq \min\Big(  (I_{1}-1)J,~ I_2K  \Big)
	\end{align}
	where $I_1\geq I_2$ and$I_2= I+1- I_{1}$ are chosen from
	\begin{align}
	\{I_1,I_2\} = \arg\max_{\{I_1,I_2\}\in\mathbb{Z}^+}~ &\min\Big((I_{1}-1)J, I_2K\Big)
	\end{align}
	Then $\A$, $\B$ and $\C$ are identifiable up to permutation and scaling of columns, almost surely.
\end{theorem}

The above method can be directly applied to $\Z$ in \eqref{Z}, if we treat $\A_r$, $\Q^T\A_t^*$, $\B$ as $\A$, $\B$, $\C$, respectively, and the corresponding dimensions are $I=M_r$, $J=N/2$ and $K=4$.
It is clear that the $\A_r$, $\B$, and $\Q^T\A_t^*$ are identifiable up to scaling and permutation ambiguities under quite mild conditions---both $N$ and $M_r$ can be smaller than the number of paths.
We remark that  $\A_r$, $\B$, and $\Q^T\A_t^*$ can be identified using any tensor factorization method, e.g., the least squares fitting formulation for tensor decomposition and ALS, which may have some other benefits such as being more noise-robust. 
Nevertheless, the employed approach admits by far the strongest identifiability result for a third-order tensor which has a Vandermonde latent factor.
Another good feature of the employed approach is that it is very lightweight and consists of only simple algebraic procedures and eigen-decomposition, which is very friendly to real-time implementation.

\subsubsection{Identification of DODs}

By the proposed procedure (or any other tensor factorization algorithm), $\A_r,\B$ and $\Q^T\A_t^*$ can be identified. However, whether $\A_t$ is identifiable from $\Q^T\A_t^*$ is not yet clear. 
Recall that we previously defined $\E=\Q^H\A_t$ in \eqref{E}. Since there is complex scaling ambiguity in the estimate of $\E$, i.e., $\hat{\E}$, the problem is equivalent to solving
$${\hat{\e}} =\xi {\Q}^H{\a_t}$$
when $\Q^H$ is a known compression (fat) matrix and $\hat{\e}$ is a given compressed measurement vector, where $\hat{\e}$ can represent any column of the estimated $\E$ and $\a_t$ represents the corresponding column in $\A_t$.
Here, $\xi$ is a complex-valued non-zero scalar that represents the scaling ambiguity inherited from the tensor factorization phase.
Solving the above underdetermined system of equations to recover the vector $\a_t$ is quite similar to the problem of \textit{compressive sensing} \cite{candes}.
However, our $\a_t$ is not sparse, and sparsity is what modern compressive sensing relies on to establish signal identifiability---this raises the question if $\a_t$ is still identifiable from the system of linear equations, since an underdetermined system could have an infinite number of solutions?

To address this issue, we have the following lemma:
\begin{lemma}\label{lem:ident}
	Given a system of equations $\hat{\e}=\xi \Q^H\a_t$ where $\Q\in\bC^{M_t\times N/2}$, $\xi\in\mathbb{C}$, $\xi \neq 0$, and $\a_t\in\bC^{M_t}$ is a function of $\theta$. Assume that $\Q$ is generated following some absolutely continuous distribution, and that $\a_t$ is a steering vector keeping the transmit array structure. Also assume that $M_t\geq N/2\geq 2$. Then, $\a_t$ is identifiable from $\hat{\e} = \xi \Q^H\a_t$ almost surely.
\end{lemma}
\begin{IEEEproof}
	Let us assume that there exists another steering vector $\a_t'$ that also satisfies $\c=\xi' \Q^H\a_t'$, where $\a_t'\neq \a_t$ and $\xi'\neq \xi$. Hence, we have
	\begin{equation}\label{eq:contradict1}
	\Q^H[ \xi\a_t,- \xi'\a_t']={\bf 0}.
	\end{equation}
	Since $\a_t$ and $\a_t'$ are Kronecker product of two Vandermonde vectors, we have
	$${\rm rank}([ \xi\a_t,- \xi'\a_t'])=2$$
	if $M_t\geq 2$.
	Also because $\Q$ is a random matrix generated following some absolutely continuous distribution and $\xi, \xi' \neq 0$, we have
	$$ {\rm rank}( \Q^H[ \xi\a_t,- \xi'\a_t']) = 2$$
	holds with probability 1 (Lemma 1, \cite{sid2012paracom})---which is a contradiction to \eqref{eq:contradict1}. This completes the proof.	
\end{IEEEproof}

Lemma~\ref{lem:ident} clearly indicates that if $M_t,N/2\geq 2$, the solution to $\hat{\E}=\Q^H\A_t\diag(\bxi) $ is unique (where $\bxi$ contains the inherited scaling ambiguities from the tensor factorization stage)---if the columns of $\A_t$ are Vandermonde vectors that have different generators. To be specific, we have the following corollary:

\begin{corollary}\label{cor:QA}
	Assume that $M_t,N/2\geq 2$, and that $\A_t$ is the manifold of an URA or ULA, which has a set of different DODs, i.e, $(\vartheta_{k},\varphi_{k})\neq (\vartheta_{j},\varphi_{j})$ for $k\neq j$. Then, $\A_t$ can be uniquely identified from the system
	$\hat{\E}=\Q^H\A_t\diag(\bxi)$.
\end{corollary}

Corollary~\ref{cor:QA} indicates that under the premise that the third-order tensor $\Z$ is identifiable, then one can use a pilot matrix has as few as $4$ columns, which can be rather economical in practice. On the other hand, the results in Lemma~\ref{lem:ident} and Corollary~\ref{cor:QA} are not entirely surprising: After all, all the columns in $\A_t$ are parametrized by only two variables---it makes much sense that we can identify them from two equations.

Combining the above results, we have the following theorem that states an integrated result of the two steps (i.e., tensor factorization and $\A_t$ identification):
\begin{theorem}\label{thm:hidden}
	Assume that the receive antenna array is a ULA and the transmit array is a URA, and that every component in $(\theta_k,\vartheta_k,\varphi_k)$ is different. Moreover, the path-loss matrix $\B$ is random.
	Then, the array manifolds $\A_r$, $\A_x$, $\A_y$ and the path-losses are uniquely identifiable with probability one via the proposed approach if
		\begin{align}\label{eq:cond2}
			K \leq \min\Big(    4(P_{r}-1), ~   \left( M_r+1- P_{r}\right)N/2   \Big)
		\end{align}
		where $P_r$ is chosen from
		\begin{align}\label{tmax}
		P_r = \arg\max_{\{t,P_r\}\in\mathbb{Z}^+}~ &t \notag\\
								\text{s. t.}~ & 4(P_{r}-1)\geq t \notag\\
											& \left( M_r+1- P_{r}\right)N/2 \geq t.
		\end{align}
\end{theorem}

The above theorem shows that when $M_t>N/2$, the identifiability is only determined by $M_r$ and $N$. This means that no matter what the transmit antenna is, as long as the number of multipaths satisfies \eqref{eq:cond2}, we can always identify the whole channel matrix. The associated algorithm, dubbed \textit{CTD}, that ensures the above identifiability result is summarized in Algorithm \ref{CTD}.

To recover $(\vartheta_{k},\varphi_{k})$ for all $k$ from $\hat{\E}=\Q^H\A_t\diag\(\bxi\)$, one can formulate this problem as a fitting problem, i.e.,
	\begin{align}\label{eq:recov}
	\min_{\vartheta_{k},\varphi_{k},\xi }&~\left\|[\hat{\E}]_{:,k} -\xi\Q^H\a_{t,k} \right\|_2^2,~\forall k=1,\cdots,K
	\end{align}
where $\a_{t,k}=\a_{y,k}\otimes\a_{x,k}$ if a URA is used at the transmitter.
Problem~\eqref{eq:recov} can be solved via many nonlinear programming algorithms since it is continuously differentiable. 
The only difficulty might be that the gradient w.r.t. $(\vartheta_{k},\varphi_{k})$ could be tedious to derive. Here, we use a simpler approximate algorithm to estimate $(\vartheta_{k},\varphi_{k})$ from $\hat{\E}=\Q^H\A_t\diag\(\bxi\)$, which works well in practice. The detailed method is presented in Appendix \ref{gradient}.

\subsubsection{Complexity Analysis}
The computational complexity for CTD consists of two parts, i.e., the smoothed ESPRIT in Step 3 and the refinement in Step 5 of Algorithm \ref{CTD}. The complexity for smoothed ESPRIT is dominated by the singular value decomposition (SVD) of $\X_s$ in Step 2, which needs $\mathcal{O}(2P_rQ_rNK+8P_r^2Q_rN+64P_r^3)$ flops. The refinement takes $\mathcal{O}(K^3+K^2NM_r+2KNM_r)$ flops. 
The overall complexity of CTD is $\mathcal{O}(2P_rQ_rNK+8P_r^2Q_rN+64P_r^3)$ flops when $K\leq\min(4(P_r-1),Q_rN/2)$. Otherwise, the complexity is $\mathcal{O}(2P_rQ_rNK+8P_r^2Q_rN+64P_r^3 + K^3+K^2NM_r+2KNM_r)$ flops.

\begin{algorithm}
	\caption{CTD for channel estimation with frugal pilot}
	\label{CTD}
	\begin{algorithmic}[1]
		\State Determine $P_r$ and the maximum $K$ from Theorem \ref{thm:hidden}, and then compute $Q_r=M_r+1-P_r$
		\State Follow \eqref{smoothedX} to construct
		 $\X_s=\(\A_{r,P_r}\odot\B\) \( \A_{r,Q_r}\odot  (\A_t^H\Q)^T\)^T $, and estimate the signal subspace $\U_s$ via the SVD of $\X_s$.
		 
%
%

       \State Apply smoothed ESPRIT \cite{smoothESPRIT} to estimate $\A_r$, $\Q^H\A_t$ and $\B$.
		\State Estimate $\theta_{k}$ from $\A_r$ and $\vartheta_{k}$ and $\varphi_{k}$ from the estimate of $\A_t^H\Q$ column-by-column via gradient descent (see Appendix \ref{gradient})
		
		\State Finally, refine $\hat{\A}_r$ from $\{\hat{\theta_{k}}\}$, $\hat{\A}_t$ from $\{\hat{\vartheta}_{k},\hat{\varphi}_k\}$ and $\hat{\B}=\Big( \big(\hat{\A}_r\odot(\Q^T\hat{\A}_t^\ast)\big)^\dagger\Z \Big)^T$

		\State Recover the channel from $\hat\A_r,\hat\A_x,\hat\A_y$ and $\hat{\B}$.
	\end{algorithmic}\label{algorithm3}
\end{algorithm}

\section{Numerical Results}
We consider a massive MIMO system with a DP URA at the BS and a DP ULA at the MS. This particular case is of considerable practical interest in 3GPP as a candidate for implementation \cite{3gpp}. 
In the simulation, we assume that the multipath propagation gains are Rician distributed, and all the multipath parameters are randomly (uniformly) drawn. The BS covers $[0^\circ,90^\circ]$ elevation angular range and $(-45^\circ,45^\circ)$ azimuth angular range, while the MS only covers $[-60^\circ,60^\circ]$ azimuth angular range since the elevation angle is zero for ULA, i.e., $\theta_k\sim \mathcal{U}(-\pi/3,\pi/3),\,
\varphi_k\sim\mathcal{U}(0,\pi/2),\,  \vartheta_k\sim\mathcal{U}(-\pi/3,\pi/3)$. 
Moreover, similar to \cite{polarization5,kappa}, we set $\kappa=13.2$ dB, which is representative of urban propagation scenarios.
We use the LS estimator of $\H$ as a baseline to compare with the reconstructed channel from the estimated key parameters, when LS is applicable. 
All the results are averaged over 500 Monte-Carlo trials using a computer with 3.2 GHz Intel Core i5-4460 and 4 GB RAM.
The normalized MSE (NMSE) of channel estimates is computed from
\begin{align}\label{nmse}
\mathrm{NMSE} = \frac{1}{500}\sum_{i=1}^{500}\|\hat{\H}_{i}-\H\|_F^2/\|\H\|_F^2
\end{align}
where $\hat{\H}_{i}$ denotes the channel that is reconstructed from the estimated key parameters from the $i$th Monte-Carlo trial.
In all the simulations, we assume that the number of paths (i.e., $K$) is known or has been estimated. Estimating $K$ is a tensor rank estimation problem, which is known to be NP-hard \cite{hillar2013most}. In practice, we are interested in the useful signal rank, i.e., the number of significant paths, and for this task we have a few practically effective algorithms, such as the one in \cite{bro2003new,da2008robust,da2011multi}.


In the first example, we compare the performance of the proposed tensor factorization-based method (implemented using alternating least squares (ALS); labeled as `PARAFAC') with IMDF \cite{liu2}, LS, multidimensional unitary ESPRIT (U-ESPRIT) \cite{unitaryesprit} and a CS
based technique \cite{jarvis} which is briefly summarized in Section II-C. Note that in the CS method, we quantize each angle using 7 bits , so the resulting dictionary has size $4M_rM_t\times 2^{23}$, which is intractable in a conventional desktop computer. To circumvent this, we implement the CS algorithm as follows. After obtaining the LS channel estimate, we first reshape each sub-block of the channel estimate as a $M_r\times M_x\times M_y$ tensor, and average the resulting tensors. Then we implement 3-D FFT with $128 \times 128 \times 128$ points to estimate $\{\theta,\vartheta,\varphi\}$, using the so-called peak-picking technique. Finally, we update the path-loss matrix $\B$ via \eqref{sp4}. 
We consider a DP MIMO scenario where the receiver has a ULA with $M_r=2$ sensors and the transmitter has a $4\times 8$ URA. Thus, the channel matrix has size $4\times 64$. The number of multipaths randomly varies from 1 to 6. A row-orthogonal pilot matrix $\S$ is used in this case. Thus, PARAFAC, IMDF and CS are performed based on the LS channel estimate. It is worth noting that we initialize PARAFAC using the IMDF estimates. Specifically, we first implement IMDF to estimate $\{\omega_{r,k}, \omega_{x,k}, \omega_{y,k}\}_{k=1}^K$ which are then used to initialize $\A_r$, $\A_x$ and $\A_y$, respectively. Finally, the $\B$ matrix is refined via the LS estimate of \eqref{sp4}.
We test the performance of all the competitors under known and unknown number of multipaths. For the latter, we set $K=6$ for all the algorithms. 

\begin{figure}
	\begin{center}
		\subfigure[known $K$]{\label{fig:mseknownChannel1} \includegraphics[width=1\linewidth, trim=0 0 0 0]{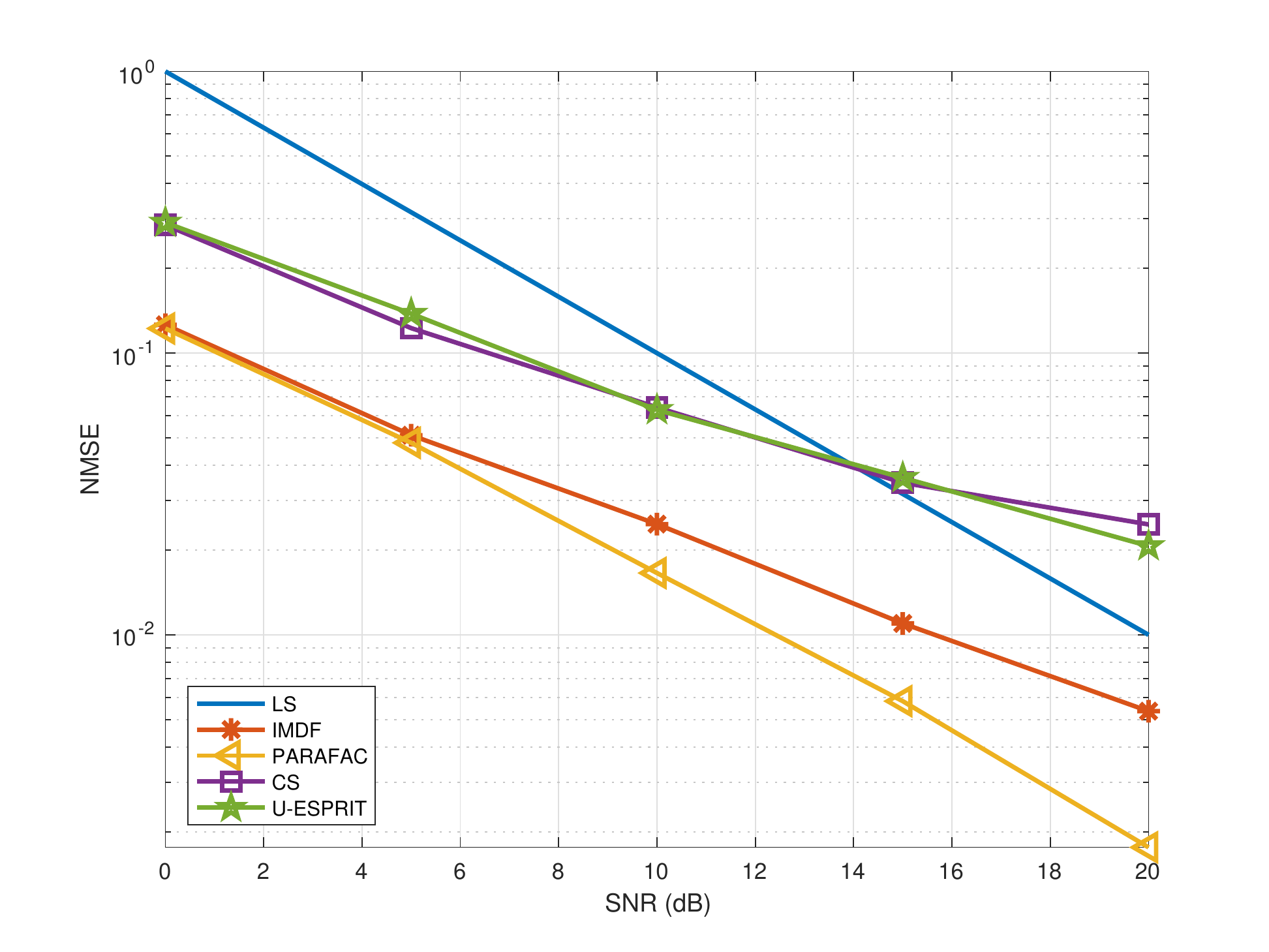}} 
		\subfigure[unknown $K$]{\label{fig:mseunkonwnChannel1} \includegraphics[width=1\linewidth, trim=0 0 0 0]{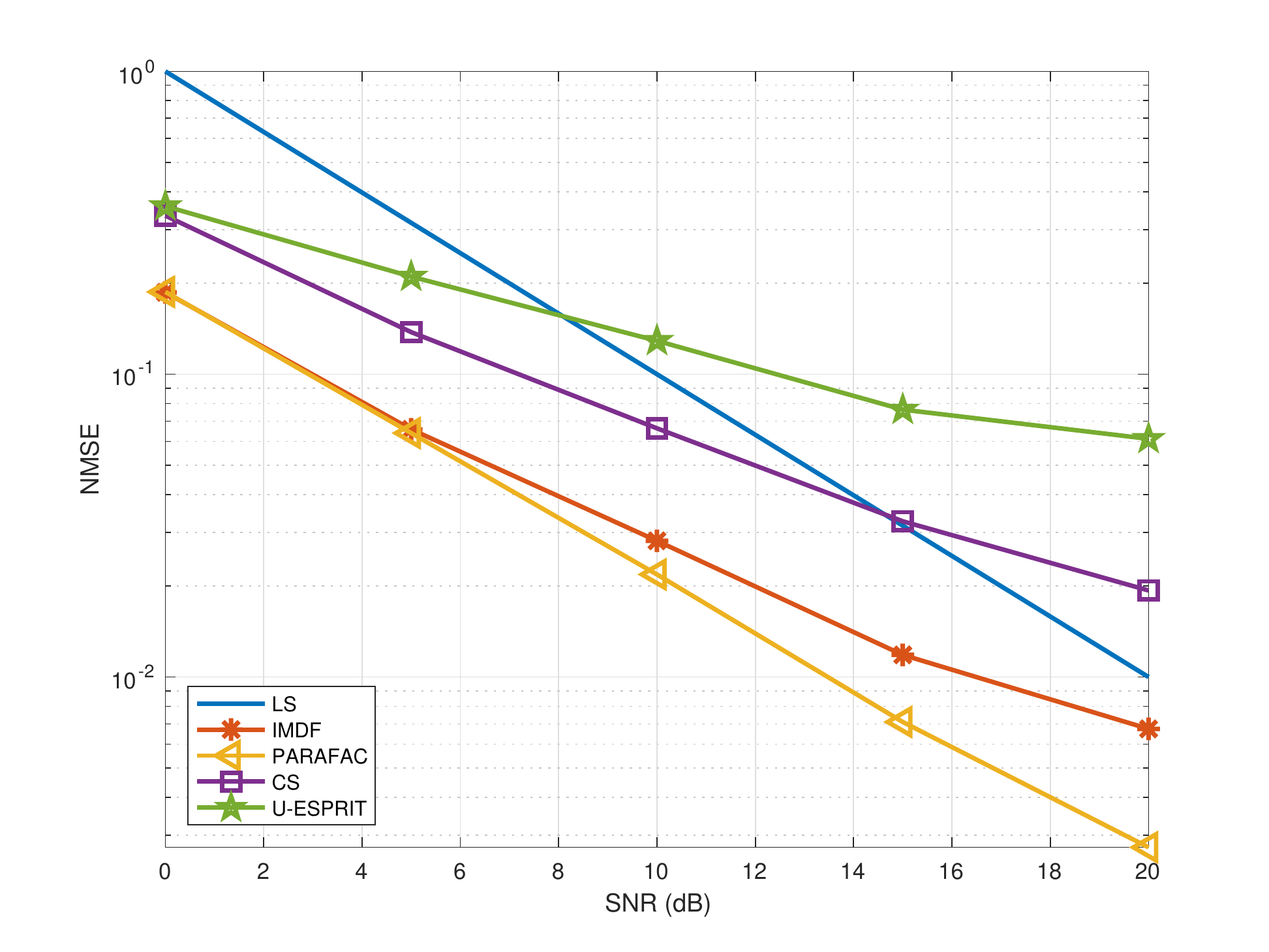}}
	\end{center}
	\caption{NMSE versus SNR.}\label{msevssnr}
\end{figure}

It is observed from Fig. \ref{msevssnr} that PARAFAC (i.e., ALS) outperforms the IMDF, U-ESPRIT, LS and CS algorithms in both cases. Compared to Fig. \ref{fig:mseknownChannel1}, PARAFAC, IMDF, U-ESPRIT and CS exhibit slight performance loss in Fig. \ref{fig:mseunkonwnChannel1}, where the exact number of multipath is unknown. When SNR $>14$ dB, we see that the NMSE of CS is even worse than the LS method. This is mainly because as SNR increases, the performance of CS is limited by the resolution of the dictionary grids. 
We observe that U-ESPRIT occasionally fail to produce reasonable results. Such failure does not occur frequently, but due to this reason the NMSE of U-ESPRIT is relatively high even in the high SNR cases. It is worth noting that if we remove such outlying cases, then U-ESPRIT performs better than the CS method but is still not as good as IMDF and ALS based PARAFAC.

\begin{figure}
	\begin{center}
		\subfigure[known $K$]{\label{fig:mseknownChannel1Mt} \includegraphics[width=1\linewidth, trim=0 0 0 0]{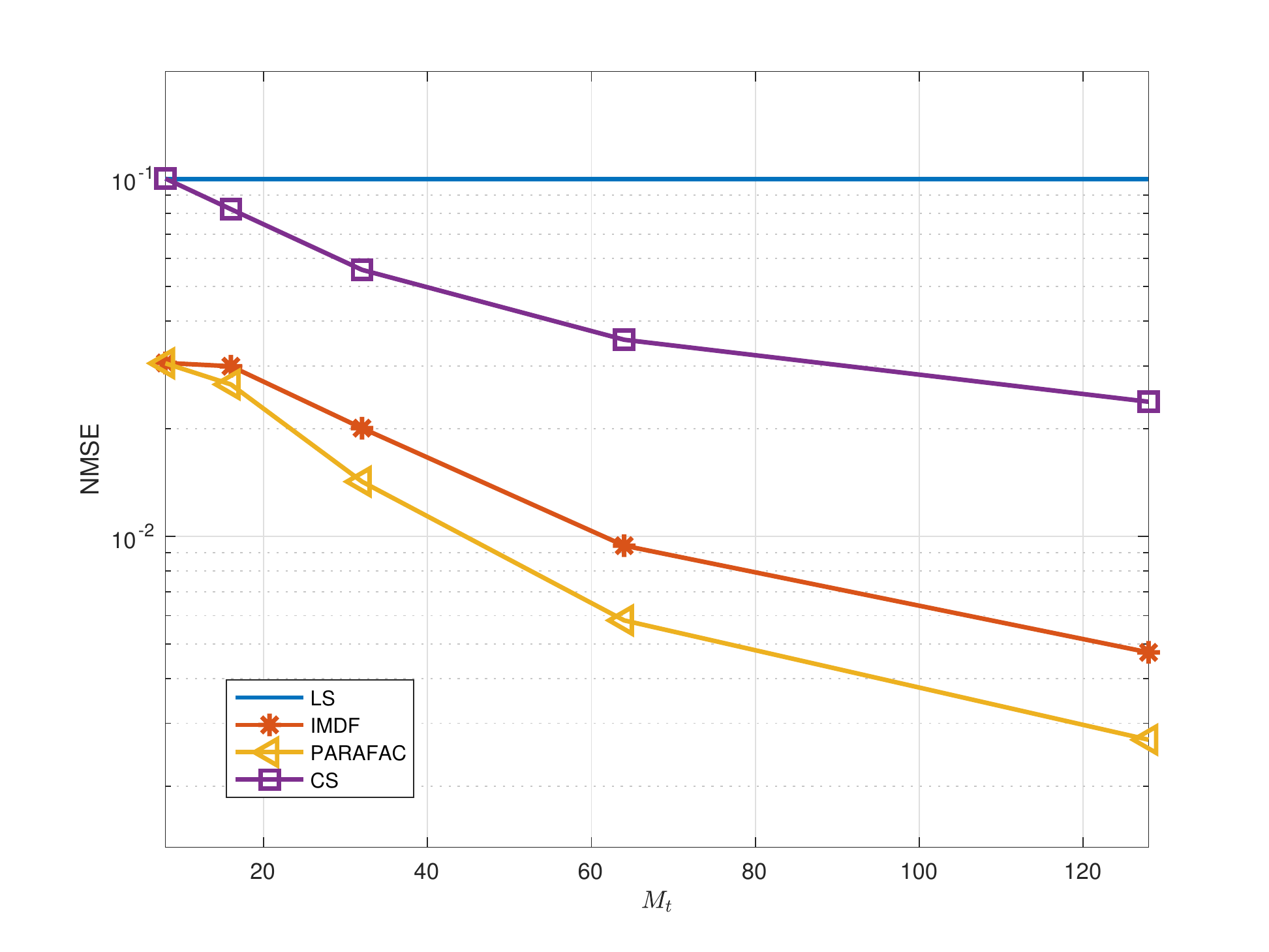}} 
		\subfigure[unknown $K$]{\label{fig:mseunkonwnChannel1Mt} \includegraphics[width=1\linewidth, trim=0 0 0 0]{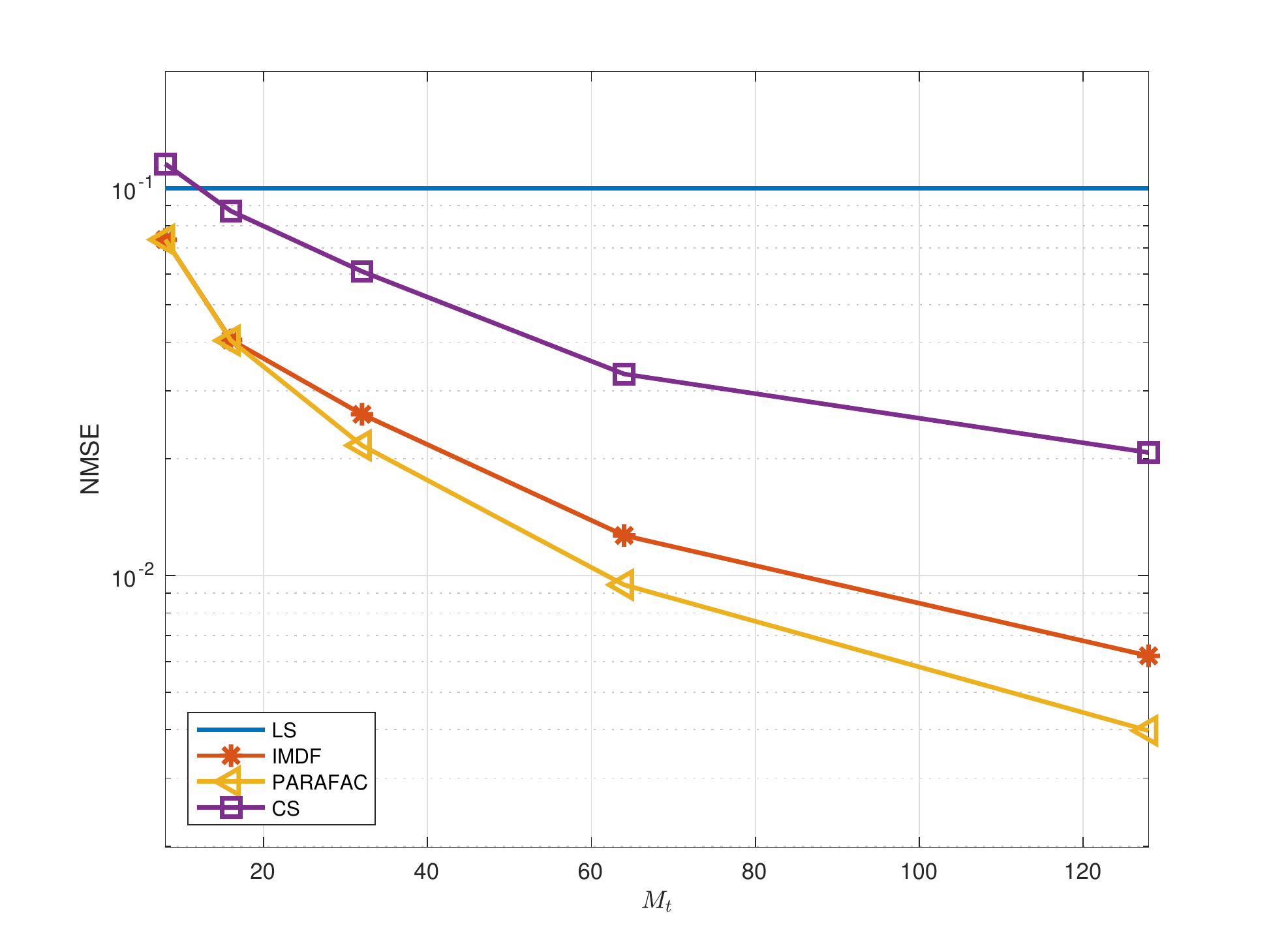}}
	\end{center}
	\caption{NMSE versus $M_t$. ($M_r=2$, SNR $=10$ dB, the number of multipaths varies from 1 to 6.)}\label{msevssnrMt}
\end{figure}

The second example examines the NMSE performance versus $M_t$, where $M_t$ varies from 8 to 128 and the size of URA with $(M_x,M_y)\in\{(2,4),(4,4),(4,8),(8,8),(8,16)\}$. SNR is fixed at 10 dB, and the other parameters keep the same as the previous example. Fig. \ref{msevssnrMt} shows similar results as Fig. \ref{msevssnr}, where PARAFAC performs the best. Combining the results in Figs. \ref{msevssnr} and \ref{msevssnrMt}, we see that  PARAFAC has similar accuracy as IMDF in low SNR and small sample size cases. With increasing SNR or $M_t$, PARAFAC outperforms IMDF evidently. 
Overall, PARAFAC performs better than IMDF. This is because PARAFAC employs the LS criterion, which corresponds to Vandermonde structure-agnostic ML for Gaussian noise, and thus is more robust to noise. IMDF is an algebraic closed-form method, which exploits Vandermonde structure and is much faster than PARAFAC, but is not optimal in handling Gaussian noise. On the other hand, PARAFAC accounts for the low-rank structure of the channel matrix, so it works well even though there are path-losses smaller than the noise variance. However, IMDF is inherently a subspace method, which picks the signal subspace according to the principal eigenvalues or singular values. Once the noise variance is greater than the path-loss values, the signal subspace may be miss-estimated. 

We now consider a DP MIMO system where $M_x=M_y=8$, $M_r=3$ and $N=16$. Thus, the task here is to recover a $6\times 128$ matrix from $6\times 16$ received data matrix. We compare the CTD method with the LS and joint orthogonal matching pursuit (J-OMP) algorithm \cite{jomp}. Specifically, for the J-OMP method, we implement it for each column of $\Z$ in \eqref{Z}, and solve the following optimization problem $\min_{\g_{i}}~ \|[\Z]_{:,i}-\bPh_\text{J-OMP}\g_{i}\|_2^2, ~\text{s. t.}~ \text{supp}(\g_{i})\leq K$,
where $\bPh_\text{J-OMP}=\((\bPh_y\otimes\bPh_x)^H\Q\)^T\otimes\bPh_r$. Finally, we use the estimates of $\{\g_{i}\}_{i=1}^4$ to recover the whole channel.
For simplicity, the dictionaries $\bPh_r,\bPh_x$ and $\bPh_y$ corresponding to $\omega_r,\omega_x$ and $\omega_y$, respectively, are all computed via uniformly dividing $[-\pi,\pi]$ into 128 points, such that $\bPh_\text{J-OMP}$ is of size $16\times 2^{21}$. We set $K=6$ for CTD and J-OMP and compare the NMSE performance by varying the number of multipaths from 1 to 6 under $\text{SNR}=20$ dB. We also plot the NMSE of CTD and J-OMP with the correct number of multipaths.

\begin{figure}
	\centering
	\includegraphics[width=1\linewidth]{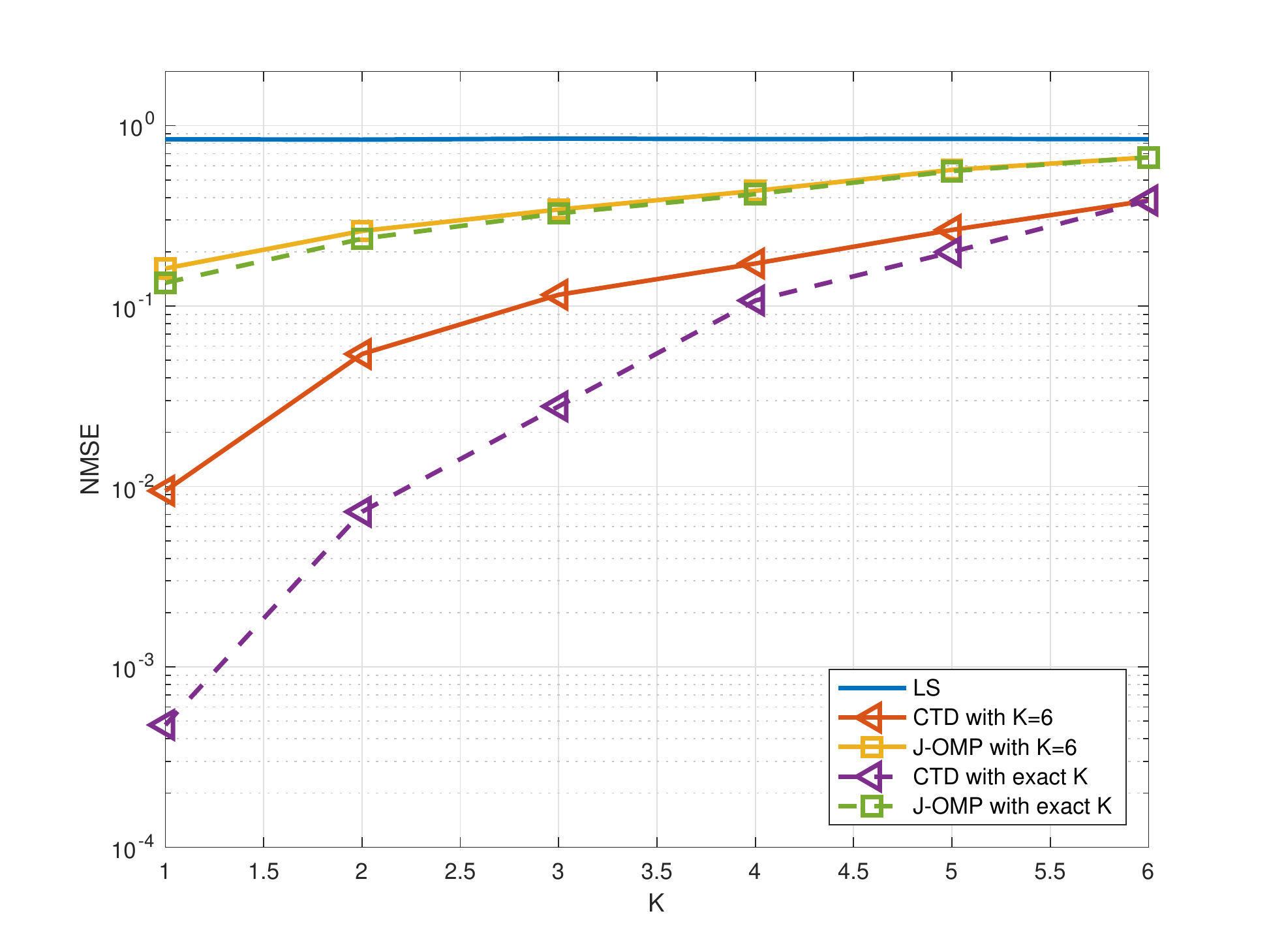}
	\caption{NMSE versus $K$.}
	\label{fig:hiddenmr3mt88n16}
\end{figure}

Fig. \ref{fig:hiddenmr3mt88n16} shows the simulation results. We see that the J-OMP and LS do not work well, while our method can offer reliable performance. The reason for the failure of LS is obvious, i.e., the LS channel estimate is rank deficient. Due to the high coherence between the grids of the flat dictionary, solving the linear inverse problem is hard, so J-OMP does not work very well. Since the CTD does not have the aforementioned issues, and its maximum number of resolvable multipaths is guaranteed by Theorem \ref{thm:hidden}, it achieves the best estimation accuracy, especially for small $K$ settings. 
The performance gap better the NMSE curves of CTD with the correct $ K< 6$ and with fixed $K=6$ is relative large when the actual number of multipaths is small. The main reason for this phenomenon is that compared to CTD with the correct $K$, the channel estimate of CTD with $K=6$ is composed of several redundant multipaths that do not appear in the actual channel. 
Furthermore, it is worth noting that the dictionary in J-OMP is huge -- it takes about 400 MB memory for storage and must be updated when $\Q$ changes. In contrast, our method is dictionary-free. In many cases, it employs only one SVD to identify the channel matrix, so that its complexity is much lower. In this example, the average CPU times for CTD and J-OMP are 0.2694 and 2.9541 seconds, respectively.

\section{Conclusion}
The downlink channel estimation problem for the DP massive MIMO system has been studied through a tensor decomposition perspective. Using row-orthogonal training pilot matrices, a low-rank tensor decomposition method was devised for channel estimation, and identifiability of the key parameters of interest was established under mild and practical conditions.
Furthermore, for non-orthogonal `frugal' training pilot matrices, a two-stage compressed tensor decomposition approach with identifiability guarantees was developed for retrieving the channel matrix and estimating the key parameters, with which the downlink training overhead can be reduced significantly. Numerical simulations support our analysis and show that the proposed schemes are very effective and promising.


\appendices
\section{}
The loading matrices $\A_x$, $\A_y$, $\A_t$ and $\B$ are uniquely identifiable up to scaling and a common column permutation ambiguity provided that
\begin{align}
\min{(M_r,K)} + \min(M_x,K) &+ \min(M_y,K) \notag\\
& + \min{(4,K)} \geq 2K + 3.
\end{align}
Specifically, if $\tilde\A_r$, $\tilde\A_x$, $\tilde\A_y$ and $\tilde{\B}$ generate $\H$ (i.e., $\breve{\H} = (\tilde\A_y^*\odot\tilde\A_x^*\odot\tilde\A_r)\tilde{\B}^T $), then it must hold that
\begin{align}
&\tilde\A_r = \A_r\bPi\bSi_1,\; \tilde\A_x = \A_x\bPi\bSi_2,\notag\\
& \tilde\A_y = \A_y\bPi\bSi_3,\; \tilde{\B} = \B\bPi\bSi_4
\end{align}
where $\bPi$ is the permutation matrix and $\{\bSi_i\}_{i=1}^4$ are diagonal scaling matrices satisfying $\bSi_1\bSi_2\bSi_3\bSi_4 = \I_K$.

\section{Updating $\vartheta_k$ and $\varphi_k$}\label{gradient}
The goal here is to estimate $\vartheta_k$ and $\varphi_k$ that satisfies the following nonlinear equations"
$$[\hat{\E}]_{:,k}=\xi_k\Q^H\a_{t,k},~\forall k=1,\cdots,K.$$
For notational convenience, let $\hat{\e}_k = [\hat{\E}]_{:,k}$ and $\v_k = \Q^H\a_{t,k}$. Solving the above is similar to the 2-D single-tone harmonic retrieval problem. However, the difficult here is that the measurement is compressed by a fat matrix, and thus the conventional harmonic retrieval methods do not apply. We note the fact that $\e_k$ is parallel to $\v_k$, and its orthogonal projection, i.e., $\P_{\e_k}^\perp = \I_{N/2} - \e_k\e_k^\dagger$, is perpendicular to $\v_k$, i.e.,
\begin{align}
\P_{\e_k}^\perp(\xi_k\v_k) = \P_{\e_k}^\perp\v_k = 0.
\end{align}
which implies that the problem of estimating $\{\vartheta_k,\varphi_k\}$ is independent of the scaling $\xi_k$. 
Therefore, we propose to solve
\begin{align}\label{2dmusic}
\min_{\vartheta_{k},\varphi_{k}}&~\left\{f=\left\|[ \P_{\e_k}^\perp\v_k \right\|_2^2\right\},~\forall k=1,\cdots,K.
\end{align}
It is easy to check that the nullity of $\P_{\e_k}^\perp$ is one, so the recovery of $\v_k$ from \eqref{2dmusic} is unique.
This way, we get rid of $\xi$ in our problem formulation.

The last step is to design an efficient method to estimate $(\vartheta_k,\varphi_k)$.
Problem \eqref{2dmusic} is a non-constrained optimization problem with a smooth objective, and thus it can be handled by gradient descent. 
To further simplify the procedure, consider the following method:
Since $\vartheta_k$ and $\varphi_k$ are embedded in the $\omega_{x,k}$ and $\omega_{y,k}$, instead of directly estimating $\vartheta_k$ and $\varphi_k$, it is easier to find $\omega_{x,k}$ and $\omega_{y,k}$ first and then update $\vartheta_k$ and $\varphi_k$ via \eqref{vartheta} and \eqref{varphi}, respectively.

To this end, let us compute the gradient w.r.t. $\omega_{x,k}$ and $\omega_{y,k}$, which equals to $\nabla f = \[ \frac{\partial f}{\partial \omega_{x,k}}~ \frac{\partial f}{\partial \omega_{y,k}} \]^T$, where
\begin{align}
\!\!\!\frac{\partial f}{\partial \omega_{x,k}} &= 2\Re\( \a_{t,k}^H\Q\P_{\e_k}^\perp\Q^H(\a_{y,k}\otimes(\t_x\circledast\a_{x,k})) \)\\
\!\!\!\frac{\partial f}{\partial \omega_{y,k}} &= 2\Re\( \a_{t,k}^H\Q\P_{\e_k}^\perp\Q^H((\t_y\circledast\a_{y,k})\otimes\a_{x,k}) \)
\end{align}
with $\t_x = j\[0\ 1\ \cdots\ M_x-1\]^T$ and $\t_y = j\[0\ 1\ \cdots\ M_y-1\]^T$.
Then we update $\omega_{x,k}$ and $\omega_{y,k}$ through
\begin{align}
\begin{bmatrix}
\omega_{x,k} \\
\omega_{y,k}
\end{bmatrix}^{(r+1)} 
= 
\begin{bmatrix}
\omega_{x,k} \\
\omega_{y,k}
\end{bmatrix}^{(r)} 
- 
\mu^{(r)}
\begin{bmatrix}
\frac{\partial f}{\partial \omega_{x,k}} \\ \frac{\partial f}{\partial \omega_{y,k}}
\end{bmatrix}^{(r)}.
\end{align}

We see that the objective in Problem \eqref{2dmusic} is the same as that of the 2-D MUSIC algorithm. Therefore, the initial point can be estimated from the following spectrum
\begin{align}\label{P}
\!\!\!\!P = 
\frac{1}{(\a_{y}(\omega_y)\otimes\a_x(\omega_x))^H\Q\P_{\e_k}^{\perp}\Q^H(\a_{y}(\omega_y)\otimes\a_x(\omega_x))}
\end{align}
where the maximum is attained at $(\a_x = \a_{x,k}, \a_y = \a_{y,k})$. This way, searching $\omega_{x}$ and $\omega_{y}$ over a certain angle range and picking the one that maximizes $P$, we can obtain the initial estimate of $\omega_{x,k}$ and $\omega_{y,k}$.
One key notice here is that the global optimal solution for $\omega_x$ and $\omega_{y}$ is the phase that corresponds to the peak of the main beam, where the optimization problem in the main beam is locally concave. Hence, we need to find an initial guess of $\omega_x$ and $\omega_{y}$ that is within the main beam. 
According to the antenna beamwidth (BW) formula for uniform linear array, i.e., $\mathrm{BW}=0.886\times\mathrm{Carrier~wavelenth}/\mathrm{Antenna~diameter}$, we can pre-calculate the half-power BW (HPBW) and then use it to determine the searching step-size. For a URA, the HPBW along the $x$- and $y$- aperture positions are approximate $0.886/M_x$ and $0.886/M_y$, respective. We choose the searching step-size as half of the HPBW, such that the initial guess may be found in the main beam. It is instructive for illustrating how to select the step-size. For example, when $M_x=M_y=8$, the HPBW is about 0.12 rad. So the step-size can be determined as 0.06 rad. In practice, the antenna usually covers only a certain angle range, e.g., $[-\pi/4,\pi/4]$. In such cases, the 2-D search is efficient.

When the transmitter is a ULA, the problem w.r.t. DOD is a 1-D single-tone estimation problem. Then \eqref{P} reduces to the cost function of the 1-D MUSIC. Since $\a_t$ is Vandermonde, we do not need search any more in this case. Instead, the root-MUSIC comes into play. More importantly, in the 1-D single-tone case, the theoretical variance of the estimate of root-MUSIC is approximately the Cram\'er-Rao bound \cite{rootmusic}. Therefore, we do not need the gradient update to further refine the root-MUSIC estimate.

\bibliographystyle{ieeetran}
\bibliography{IEEEabrv,mybib}

\begin{thebibliography}{10}
\providecommand{\url}[1]{#1}
\csname url@samestyle\endcsname
\providecommand{\newblock}{\relax}
\providecommand{\bibinfo}[2]{#2}
\providecommand{\BIBentrySTDinterwordspacing}{\spaceskip=0pt\relax}
\providecommand{\BIBentryALTinterwordstretchfactor}{4}
\providecommand{\BIBentryALTinterwordspacing}{\spaceskip=\fontdimen2\font plus
\BIBentryALTinterwordstretchfactor\fontdimen3\font minus
  \fontdimen4\font\relax}
\providecommand{\BIBforeignlanguage}[2]{{%
\expandafter\ifx\csname l@#1\endcsname\relax
\typeout{** WARNING: IEEEtran.bst: No hyphenation pattern has been}%
\typeout{** loaded for the language `#1'. Using the pattern for}%
\typeout{** the default language instead.}%
\else
\language=\csname l@#1\endcsname
\fi
#2}}
\providecommand{\BIBdecl}{\relax}
\BIBdecl

\bibitem{icassp}
C.~Qian, X.~Fu, N.~D. Sidiropoulos, and Y.~Yang, ``Tensor-based parameter
  estimation of double directional massive mimo channel with dual-polarized
  antennas,'' in \emph{IEEE Int. Conf. Acoust., Speech, Signal Process.
  (ICASSP)}, Calgary, Canada, April 2018.

\bibitem{polarization2}
D.~Dupleich, S.~H{\"a}fner, C.~Schneider, R.~M{\"u}ller, R.~Thom{\"a}, J.~Luo,
  N.~Iqbal, E.~Schulz, X.~Lu, and G.~Wang, ``Double-directional and
  dual-polarimetric indoor measurements at 70 ghz,'' in \emph{IEEE 26th Annual
  Int. Sym. on Personal, Indoor, and Mobile Radio Commun. (PIMRC)}, Hong Kong,
  2015, pp. 2234--2238.

\bibitem{3gpp}
Z.~Bai, ``Evolved universal terrestrial radio access (e-utra); physical layer
  procedures,'' \emph{3GPP, Sophia Antipolis, Technical Specification}, 36.213
  v. 11.4.0, 2013.

\bibitem{3gpp_stand}
A.~Kammoun, H.~Khanfir, Z.~Altman, M.~Debbah, and M.~Kamoun, ``Preliminary
  results on 3d channel modeling: From theory to standardization,''
  \emph{{IEEE} J. Sel. Areas Commun.}, vol.~32, no.~6, pp. 1219--1229, 2014.

\bibitem{polarization5}
D.~Zhu, J.~Choi, and R.~W. Heath, ``Two-dimensional aod and aoa acquisition for
  wideband millimeter-wave systems with dual-polarized mimo,'' \emph{IEEE
  Trans. Wireless Commun.}, vol.~16, no.~12, pp. 7890--7905, 2017.

\bibitem{polarization4}
G.~J. Foschini and M.~J. Gans, ``On limits of wireless communications in a
  fading environment when using multiple antennas,'' \emph{Wireless personal
  commun.}, vol.~6, no.~3, pp. 311--335, 1998.

\bibitem{limitedfeedback2}
Z.~Gao, L.~Dai, Z.~Wang, and S.~Chen, ``Spatially common sparsity based
  adaptive channel estimation and feedback for fdd massive mimo,'' \emph{{IEEE}
  Trans. Signal Process.}, vol.~63, no.~23, pp. 6169--6183, 2015.

\bibitem{jarvis}
W.~U. Bajwa, J.~Haupt, A.~M. Sayeed, and R.~Nowak, ``Compressed channel
  sensing: A new approach to estimating sparse multipath channels,''
  \emph{Proc. of the IEEE}, vol.~98, no.~6, pp. 1058--1076, 2010.

\bibitem{jomp}
X.~Rao and V.~K. Lau, ``Distributed compressive csit estimation and feedback
  for fdd multi-user massive mimo systems,'' \emph{{IEEE} Trans. Signal
  Process.}, vol.~62, no.~12, pp. 3261--3271, 2014.

\bibitem{panos}
P.~N. Alevizos, X.~Fu, N.~Sidiropoulos, Y.~Yang, and A.~Bletsas, ``Non-uniform
  directional dictionary-based limited feedback for massive mimo systems,'' in
  \emph{IEEE Int. Sym. Model. and Optimization in Mobile, Ad Hoc, and Wireless
  Networks (WiOpt)}, Paris, France, 2017, pp. 1--8.

\bibitem{fangjun}
Z.~Zhou, J.~Fang, L.~Yang, H.~Li, Z.~Chen, and R.~S. Blum, ``Low-rank tensor
  decomposition-aided channel estimation for millimeter wave mimo-ofdm
  systems,'' \emph{{IEEE} J. Sel. Areas Commun.}, vol.~35, no.~7, pp.
  1524--1538, 2017.

\bibitem{fangjun2}
Z.~Zhou, J.~Fang, L.~Yang, H.~Li, Z.~Chen, and S.~Li, ``Channel estimation for
  millimeter-wave multiuser mimo systems via parafac decomposition,''
  \emph{IEEE Trans. Wireless Commun.}, vol.~15, no.~11, pp. 7501--7516, 2016.

\bibitem{music}
R.~Schmidt, ``Multiple emitter location and signal parameter estimation,''
  \emph{{IEEE} Trans. Antennas Propag.}, vol.~34, no.~3, pp. 276--280, 1986.

\bibitem{esprit}
R.~Roy and T.~Kailath, ``Esprit-estimation of signal parameters via rotational
  invariance techniques,'' \emph{{IEEE} Trans. Acoust., Speech, Signal
  Process.}, vol.~37, no.~7, pp. 984--995, 1989.

\bibitem{Sid2017}
N.~D. Sidiropoulos, L.~De~Lathauwer, X.~Fu, K.~Huang, E.~E. Papalexakis, and
  C.~Faloutsos, ``Tensor decomposition for signal processing and machine
  learning,'' \emph{{IEEE} Trans. Signal Process.}, vol.~65, no.~13, pp.
  3551--3582, 2017.

\bibitem{polarized1}
J.~Li and R.~Compton, ``Two-dimensional angle and polarization estimation using
  the esprit algorithm,'' \emph{{IEEE} Trans. Antennas Propag.}, vol.~40,
  no.~5, pp. 550--555, 1992.

\bibitem{scme}
D.~S. Baum, J.~Hansen, and J.~Salo, ``An interim channel model for beyond-3g
  systems: extending the 3gpp spatial channel model (scm),'' in \emph{IEEE
  Vehicular Technology Conference}, vol.~5, 2005, pp. 3132--3136.

\bibitem{winner}
I.~WINNER, ``Channel models, d1. 1.2 v1. 2,'' IST-4-027756 WINNER II
  Deliverable, 4 February, Tech. Rep., 2008.

\bibitem{itu}
M.~Series, ``Guidelines for evaluation of radio interface technologies for
  imt-advanced,'' \emph{Report ITU}, vol. 638, 2009.

\bibitem{tibshirani1996regression}
R.~Tibshirani, ``Regression shrinkage and selection via the lasso,''
  \emph{Journal of the Royal Statistical Society. Series B (Methodological)},
  pp. 267--288, 1996.

\bibitem{kruskal}
J.~B. Kruskal, ``Three-way arrays: rank and uniqueness of trilinear
  decompositions, with application to arithmetic complexity and statistics,''
  \emph{Linear algebra and its applications}, vol.~18, no.~2, pp. 95--138,
  1977.

\bibitem{nikos1}
N.~D. Sidiropoulos and R.~Bro, ``On the uniqueness of multilinear decomposition
  of n-way arrays,'' \emph{Journal of Chemometrics: A Journal of the
  Chemometrics Society}, vol.~14, no.~3, pp. 229--239, 2000.

\bibitem{sid2001ident}
N.~D. Sidiropoulos and X.~Liu, ``Identifiability results for blind beamforming
  in incoherent multipath with small delay spread,'' \emph{{IEEE} Trans. Signal
  Process.}, vol.~49, no.~1, pp. 228--236, 2001.

\bibitem{parafac2}
K.~Huang, N.~D. Sidiropoulos, and A.~P. Liavas, ``A flexible and efficient
  algorithmic framework for constrained matrix and tensor factorization,''
  \emph{{IEEE} Trans. Signal Process.}, vol.~64, no.~19, pp. 5052--5065, 2016.

\bibitem{parafac3}
L.~Sorber, M.~Van~Barel, and L.~De~Lathauwer, ``Optimization-based algorithms
  for tensor decompositions: Canonical polyadic decomposition, decomposition in
  rank-($l_r,l_r,1$) terms, and a new generalization,'' \emph{SIAM Journal on
  Optimization}, vol.~23, no.~2, pp. 695--720, 2013.

\bibitem{freqest1}
C.~Qian, L.~Huang, H.-C. So, N.~D. Sidiropoulos, and J.~Xie, ``Unitary puma
  algorithm for estimating the frequency of a complex sinusoid.'' \emph{{IEEE}
  Trans. Signal Process.}, vol.~63, no.~20, pp. 5358--5368, 2015.

\bibitem{freqest2}
D.~Rife and R.~Boorstyn, ``Single tone parameter estimation from discrete-time
  observations,'' \emph{{IEEE} Trans. Inf. Theory}, vol.~20, no.~5, pp.
  591--598, 1974.

\bibitem{liu2}
J.~Liu and X.~Liu, ``An eigenvector-based approach for multidimensional
  frequency estimation with improved identifiability,'' \emph{{IEEE} Trans.
  Signal Process.}, vol.~54, no.~12, pp. 4543--4556, 2006.

\bibitem{xu2013block}
Y.~Xu and W.~Yin, ``A block coordinate descent method for regularized
  multiconvex optimization with applications to nonnegative tensor
  factorization and completion,'' \emph{SIAM Journal on imaging sciences},
  vol.~6, no.~3, pp. 1758--1789, 2013.

\bibitem{smoothESPRIT}
M.~S{\o}rensen and L.~De~Lathauwer, ``Blind signal separation via tensor
  decomposition with vandermonde factor: Canonical polyadic decomposition,''
  \emph{{IEEE} Trans. Signal Process.}, vol.~61, no.~22, pp. 5507--5519, 2013.

\bibitem{candes}
E.~J. Cand{\`e}s and M.~B. Wakin, ``An introduction to compressive sampling,''
  \emph{IEEE Signal Process. Mag.}, vol.~25, no.~2, pp. 21--30, 2008.

\bibitem{sid2012paracom}
N.~D. Sidiropoulos and A.~Kyrillidis, ``Multi-way compressed sensing for sparse
  low-rank tensors,'' \emph{{IEEE} Signal Process. Lett.}, vol.~19, no.~11, pp.
  757--760, 2012.

\bibitem{kappa}
Z.~Muhi-Eldeen, L.~Ivrissimtzis, and M.~Al-Nuaimi, ``Modelling and measurements
  of millimetre wavelength propagation in urban environments,'' \emph{IET
  Microwaves, Ant. \& prop.}, vol.~4, no.~9, pp. 1300--1309, 2010.

\bibitem{hillar2013most}
C.~J. Hillar and L.-H. Lim, ``Most tensor problems are np-hard,'' \emph{Journal
  of the ACM (JACM)}, vol.~60, no.~6, p.~45, 2013.

\bibitem{bro2003new}
R.~Bro and H.~A. Kiers, ``A new efficient method for determining the number of
  components in parafac models,'' \emph{Journal of Chemometrics: A Journal of
  the Chemometrics Society}, vol.~17, no.~5, pp. 274--286, 2003.

\bibitem{da2008robust}
J.~P.~C. da~Costa, M.~Haardt, and F.~Romer, ``Robust methods based on the hosvd
  for estimating the model order in parafac models,'' in \emph{Sensor Array and
  Multichannel Signal Processing Workshop, 2008. SAM 2008. 5th IEEE}.\hskip 1em
  plus 0.5em minus 0.4em\relax IEEE, 2008, pp. 510--514.

\bibitem{da2011multi}
J.~P. C.~L. da~Costa, F.~Roemer, M.~Haardt, and R.~T. de~Sousa,
  ``Multi-dimensional model order selection,'' \emph{EURASIP Journal on
  Advances in Signal Processing}, vol. 2011, no.~1, p.~26, 2011.

\bibitem{unitaryesprit}
M.~Haardt, F.~Roemer, and G.~Del~Galdo, ``Higher-order svd-based subspace
  estimation to improve the parameter estimation accuracy in multidimensional
  harmonic retrieval problems,'' \emph{{IEEE} Trans. Signal Process.}, vol.~56,
  no.~7, pp. 3198--3213, 2008.

\bibitem{rootmusic}
B.~D. Rao and K.~S. Hari, ``Performance analysis of root-music,'' \emph{{IEEE}
  Trans. Acoust., Speech, Signal Process.}, vol.~37, no.~12, pp. 1939--1949,
  1989.

\end{thebibliography}

\begin{IEEEbiography}
	[{\includegraphics[width=1in,height=1.25in,clip,keepaspectratio]{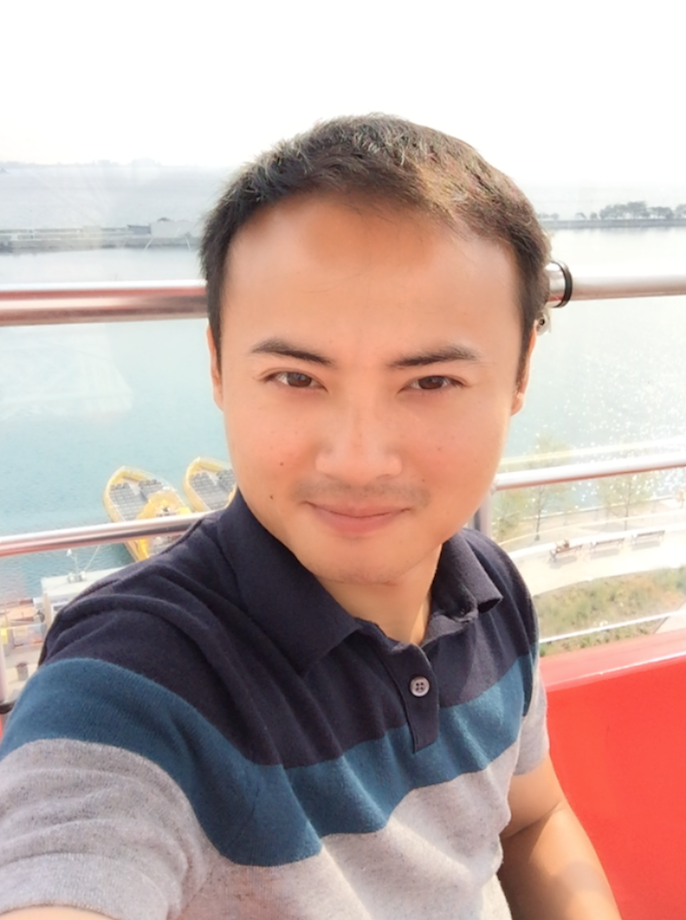}}]
	{Cheng Qian} was born in Deqing, Zhejiang, China. He received the B.E. degree in communication engineering from Hangzhou Dianzi University, Hangzhou, China, in 2011, M.S. and Ph.D. degrees in information and communication engineering from Harbin Institute of Technology, China, in 2013 and 2017, respectively. 
	He was a visiting Ph.D. student in the Department of ECE, University of Minnesota Twin Cities, Minneapolis, MN, United States, from 2014-2016. In the beginning of 2017, he returned back to the group in UMN as a postdoc for half a year. After that, he joined University of Virginia, where he is currently a postdoctoral associate in the Department of ECE.
	His research interests include tensor decomposition, optimization with applications in signal processing, wireless communications and bioinformatics.
\end{IEEEbiography}

\begin{IEEEbiography}
	[{\includegraphics[width=1in,height=1.25in,clip,keepaspectratio]{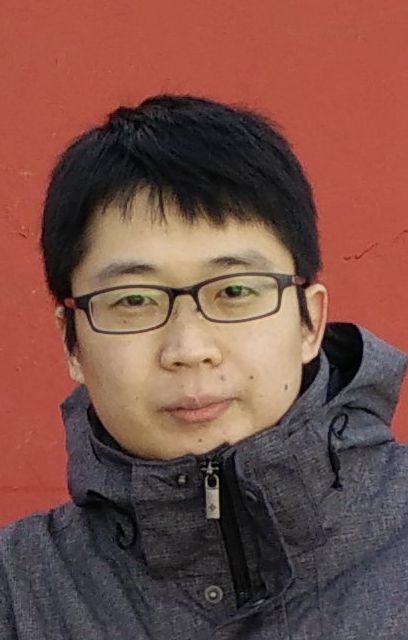}}]
	{Xiao Fu} (S'12-M'15) is an Assistant Professor in the School of Electrical Engineering and Computer Science, Oregon State University, Corvallis, OR, United States. He received his Ph.D. degree in Electronic Engineering from The Chinese University of Hong Kong (CUHK), Hong Kong, 2014. He was a Postdoctoral Associate in the Department of Electrical and Computer Engineering, University of Minnesota, Minneapolis, MN, United States, from 2014-2017. His research interests include the broad area of signal processing and machine learning, with recent emphases on tensor/matrix decomposition and multiview analysis. He received a Best Student Paper Award at ICASSP 2014, and co-authored a Best Student Paper Award at IEEE CAMSAP 2015.
\end{IEEEbiography}

\begin{IEEEbiography}	
	[{\includegraphics[width=1in,height=1.25in,clip,keepaspectratio]{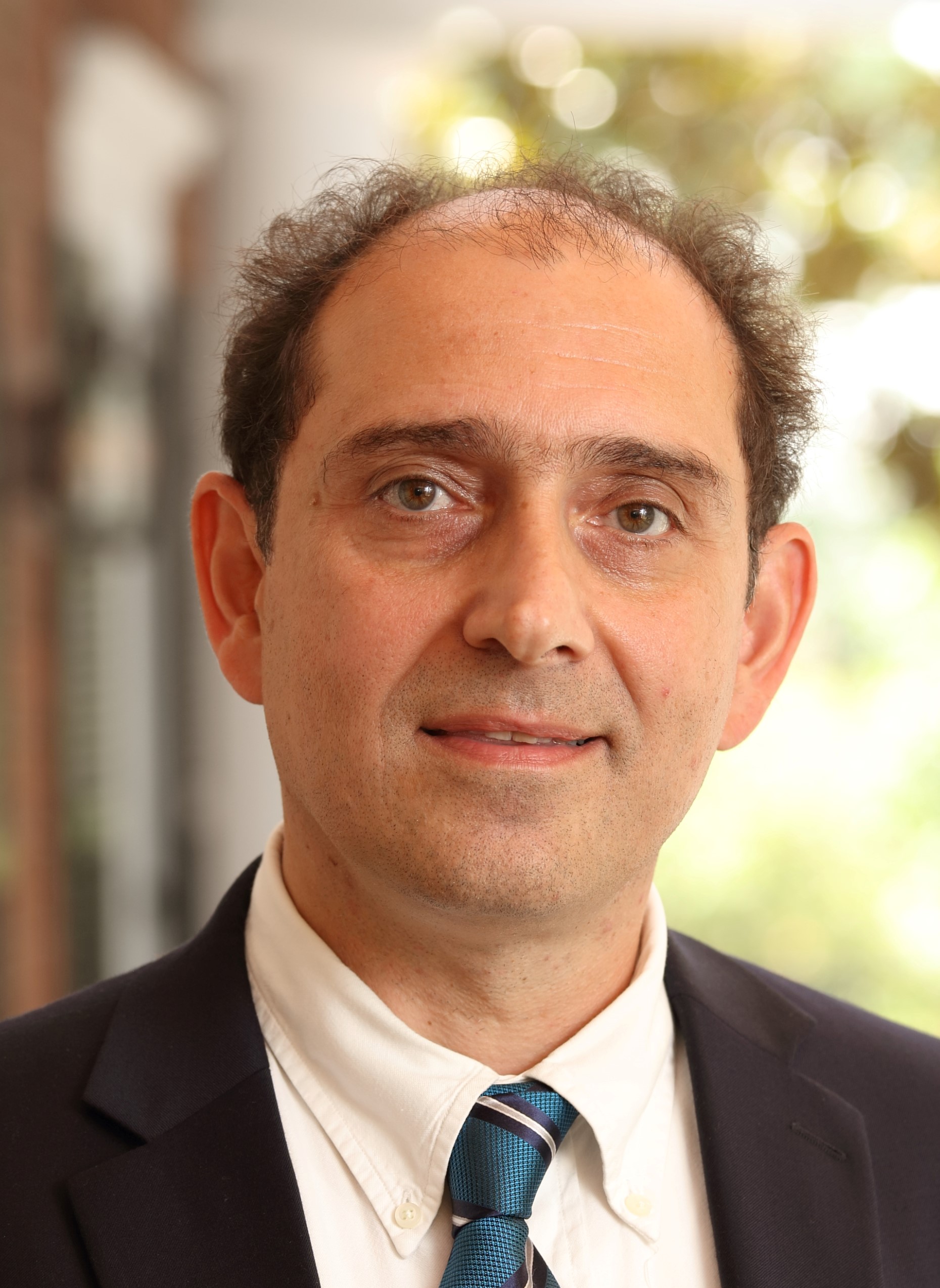}}]	
	{Nicholas D. Sidiropoulos} (F'09) earned the Diploma in Electrical Engineering from Aristotle University of Thessaloniki, Greece, and M.S. and Ph.D. degrees in Electrical Engineering from the University of Maryland at College Park, in 1988, 1990 and 1992, respectively. He has served on the faculty of the University of Virginia, University of Minnesota, and the Technical University of Crete, Greece, prior to his current appointment as Louis T. Rader Professor and Chair of ECE at UVA. From 2015 to 2017 he was an ADC Chair Professor at the University of Minnesota. His research interests are in signal processing, communications, optimization, tensor decomposition, and factor analysis, with applications in machine learning and communications. He received the NSF/CAREER award in 1998, the IEEE Signal Processing Society (SPS) Best Paper Award in 2001, 2007, and 2011, served as IEEE SPS Distinguished Lecturer (2008-2009), and currently serves as Vice President - Membership of IEEE SPS. He received the 2010 IEEE Signal Processing Society Meritorious Service Award, and the 2013 Distinguished Alumni Award from the University of Maryland, Dept. of ECE. He is a Fellow of IEEE (2009) and a Fellow of EURASIP (2014). 	
\end{IEEEbiography}

\begin{IEEEbiography}
	[{\includegraphics[width=1in,height=1.25in,clip,keepaspectratio]{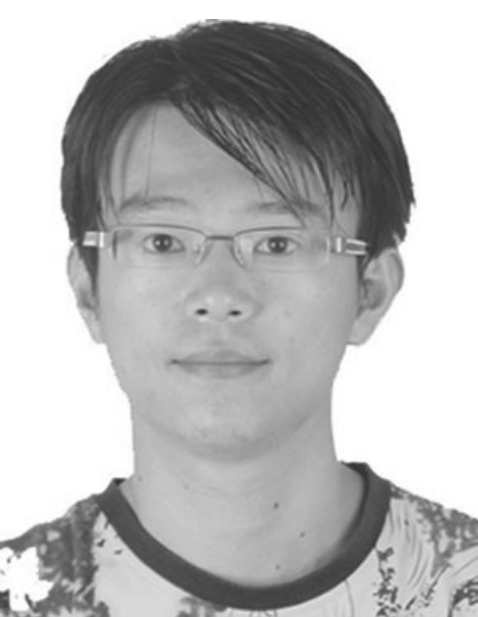}}]{Ye Yang}
	 received the B.S. degree in communications engineering, and the Ph.D. degree in communication and information systems from Xidian University, Xi’an, China, in 2008 and 2013, respectively. From January to July 2011, he was a visiting Ph.D. student with National Tsing Hua University, Hsinchu, Taiwan. From February to August 2012, he was a Visiting Scholar with the Chinese University of Hong Kong, Hong Kong. He is currently working as a Research Engineer with the RAN Algorithm Department, Huawei Technologies Investment Company, Shanghai, China. His research interests lie in algorithm design for signal processing in wireless communications with an emphasis on massive MIMO. He was the recipient of the Excellent Doctoral Dissertation Award of Shaanxi Province in 2016, and the Best Paper Award of the IEEE SIGNAL PROCESSING LETTERS 2016.
\end{IEEEbiography}

\end{document}